\documentclass[12pt]{article}
\input epsf
\usepackage{amsmath}
\usepackage{amssymb}
\usepackage{theorem}
\usepackage{graphicx}

\numberwithin{equation}{section}

\newtheorem{thm}{Theorem}[section]

{\theorembodyfont{\rmfamily}

\newtheorem{rmk}[thm]{Remark}
}

\newcommand{\qed}{\hfill \mbox{\raggedright \rule{.07in}{.1in}}}

\newcommand{\var}{\operatorname{var}}

\title{On the Implementation of the $0$--$1$ Test for Chaos}

\author{
Georg A. Gottwald
\thanks{%
Mathematics and Statistics, University of Sydney, NSW 2006, Australia.
gottwald@maths.usyd.edu.au}
\and
Ian Melbourne
\thanks{%
Mathematics and Statistics,
University of Surrey,
Guildford, Surrey GU2 7XH, UK.  ism@math.uh.edu}
}

\date{}

\begin{document}

\maketitle


\begin{abstract}
In this paper we address practical aspects of the implementation of
the $0$-$1$ test for chaos in deterministic systems. 
In addition, we present a new formulation
of the test which significantly increases its sensitivity. The test
can be viewed as a method to distill a binary quantity from the power
spectrum.  The implementation is guided by recent results from the
theoretical justification of the test as well as by exploring better
statistical methods to determine the binary quantities. We give
several examples to illustrate the improvement.
\end{abstract}


\section{Introduction} 
\label{sec-intro}
Being able to distinguish between regular and chaotic dynamics in a
deterministic system is an important question with applications ranging
from cardiac arrhythmias to the stability of our solar system. Much
progress has been made in developing tests for chaos
\cite{Kantz,Laskar93,Skokos01,Fouchard02,Cincotta03,Barrio05}.
Recently we have introduced a binary test for chaos, the $0$--$1$
test, designed for the analysis of deterministic dynamical systems
\cite{GM04,GM05}. The test distinguishes between regular and chaotic
dynamics for a deterministic system. The nature of the dynamical
system is irrelevant for the implementation of the test; it is
applicable to data generated from maps, ordinary differential
equations and partial differential equations. The test has been
applied to noisy numerical data~\cite{GM05}, experimental
data~\cite{FGMW07}, quasiperiodically forced systems and strange
nonchaotic attractors~\cite{Dawes07}, Hamiltonian
systems~\cite{Skokos04}, nonsmooth systems~\cite{Litak07} and fluid
dynamics~\cite{JulienWeiss06}.

The usual test of whether a deterministic dynamical system is chaotic
or nonchaotic involves the calculation of the maximal Lyapunov
exponent $\lambda$ \cite{Kantz}. A positive maximal Lyapunov exponent
indicates chaos: if $\lambda>0$, then nearby trajectories separate
exponentially and if $\lambda\le0$, then nearby trajectories remain in
a close neighbourhood of each other. This approach has been widely
used for dynamical systems whose equations are known. If the equations
are not known or one wishes to examine experimental data, then
$\lambda$ may be estimated using the phase space reconstruction method
of Takens \cite{Takens81}, by approximating the linearisation of the
evolution operator \cite{SanoSawada}, or by the ``direct
method''~\cite{Rosenstein93}.

In contrast our test does not depend on phase space reconstruction but
rather works directly with the time series given. The main advantages
of our test are (i) it is binary (minimizing issues of distinguishing
small positive numbers from zero), (ii) the nature of the vector field
as well as its dimensionality does not pose practical limitations, and
(iii) it does not suffer from the difficulties associated with phase
space reconstruction \cite{Kantz}. 

In this paper, we describe in detail how to implement the $0$--$1$ test for 
chaos. In addition, we carry out modifications to the test that greatly
improve the previous versions in~\cite{GM04,GM05}.

Throughout the paper, we use the logistic map to illustrate our
claims, with the exception of Section~\ref{sec-cont} where we use the
Lorenz attractor as an example of a continuous time system. The reader
can verify that our results apply equally well to other systems,
including those considered in our previous papers~\cite{GM04,GM05}.

\subsection{Recipe for the $0$--$1$ test}
\label{sec-recipe}
We briefly review how the test is implemented. Given an observation
$\phi(j)$ for $j=1,\ldots ,N$ we perform the following sequence of
steps:
\begin{enumerate}
\item For $c\in(0,\pi)$, we compute the translation variables 
\begin{align} \label{eq-p}
p_c(n)  =\sum_{j=1}^n \phi(j) \cos jc, \quad
q_c(n)  =\sum_{j=1}^n \phi(j) \sin jc
\end{align}
for $n=1,2,\ldots,N$. Typical plots of $p$ and $q$ for regular and
chaotic dynamics are given in Fig.~\ref{fig-pq}.
\begin{figure}[htb]
\centerline{%
\includegraphics[width=.4\textwidth,height=.4\textwidth]{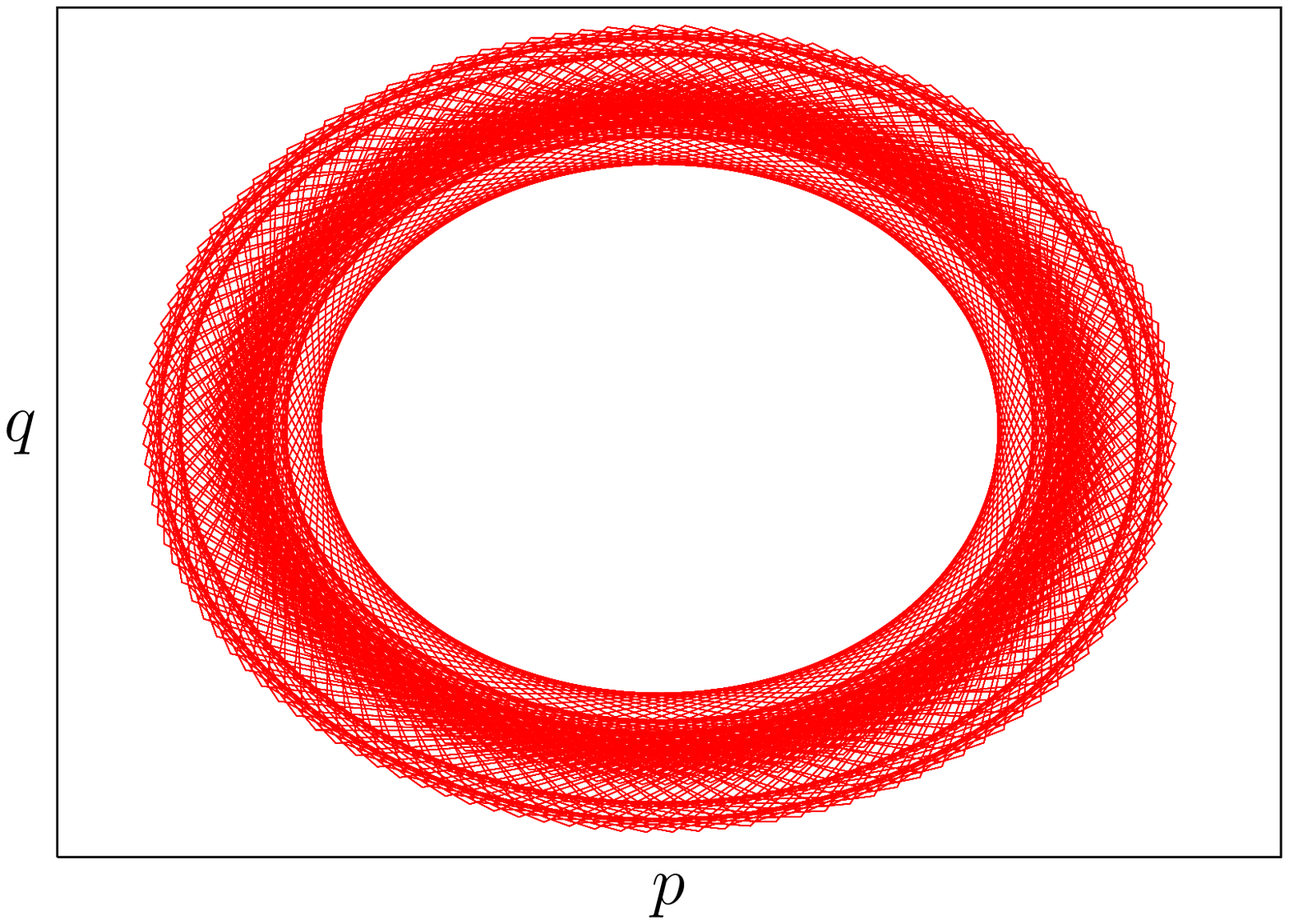}
\qquad
\includegraphics[width=.4\textwidth,height=.4\textwidth]{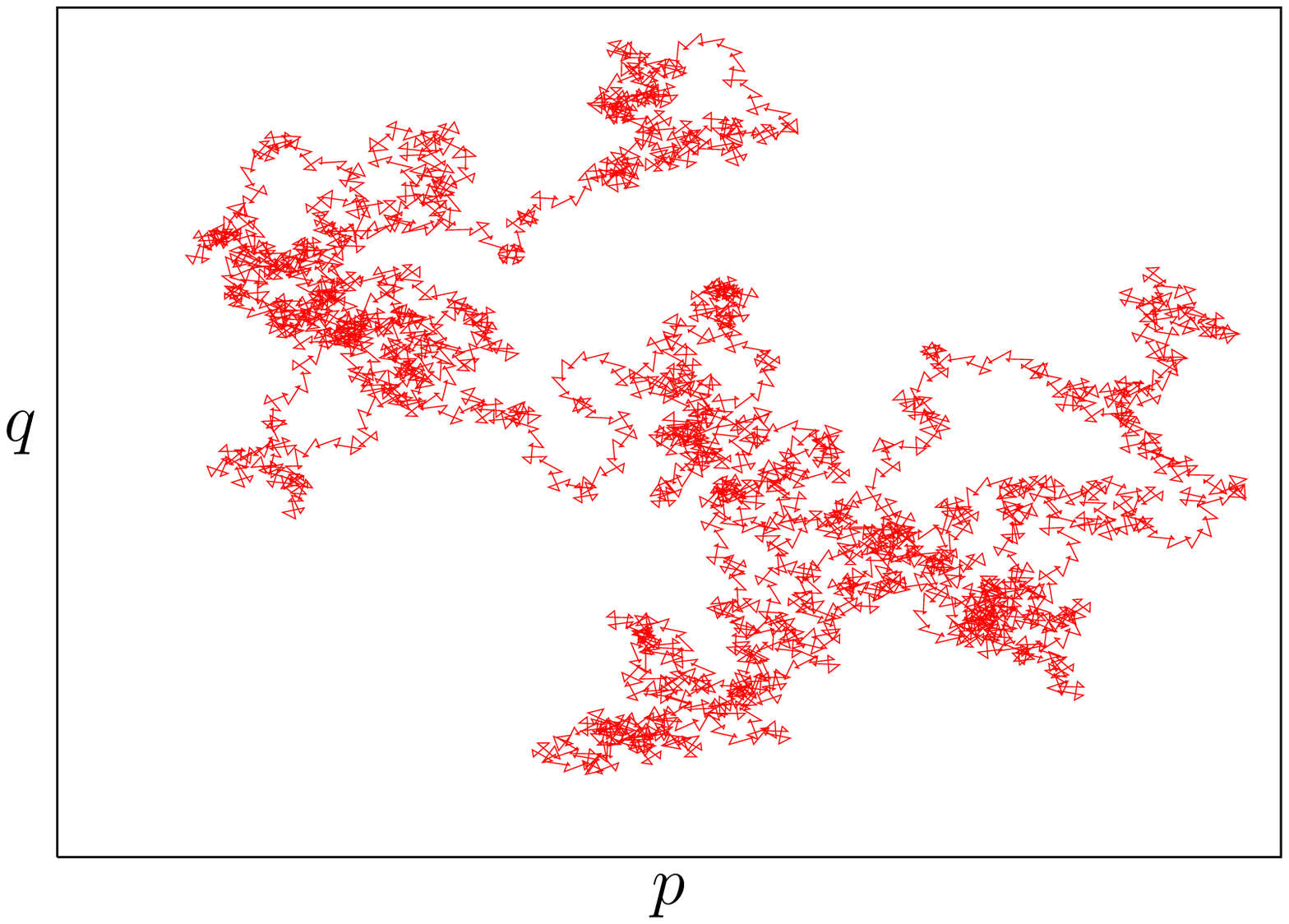}
}
\caption{
Plot of $p$ versus $q$ for the logistic map $x_{n+1}=\mu
x_n(1-x_n)$. Left: Regular dynamics at $\mu=3.55$; Right: Chaotic
dynamics at $\mu=3.9$. We used $5000$ data points. }
\label{fig-pq}
\end{figure}

\item The diffusive (or non-diffusive) behaviour of $p_c$ and $q_c$ can
be investigated by analyzing the mean square displacement
$M_c(n)$. The theory behind our test assures that if the dynamics is
regular then the mean square displacement is a bounded function in
time, whereas if the dynamics is chaotic then the mean square
displacement scales linearly with time. In Section~\ref{sec-MSQ} we
look at expressions for the mean square displacement and describe how
one may use analytical expressions derived in \cite{GMprep} to
conveniently modify the expression for the mean square displacement.

\item
We then compute the asymptotic growth rate $K_c$ of the mean square
displacement. Methods for the most effective estimation of this are
discussed in Section~\ref{sec-Kc}.

\item
Steps 1--3 are performed for $N_c$ values of $c$ chosen randomly in
the interval $(0,\pi)$. In practice, $N_c=100$ is sufficient.  The
choice of $c$ is discussed further in Section~\ref{sec-c}.  We then
compute the median of these $N_c$ values of $K_c$ to compute the final
result $K={\rm{median}}(K_c)$. Our test states that a value of
$K\approx 0$ indicates regular dynamics, and $K\approx 1$ indicates
chaotic dynamics.
\end{enumerate}

In this paper, we explore practical issues arising in the
implementation of the above algorithm. Various issues associated with
steps 2--4 are discussed in \mbox{Sections~\ref{sec-MSQ}--\ref{sec-c}}
respectively. In Section~\ref{sec-confidence} we examine finite data
size effects. In particular we look at weak chaos.  In
Section~\ref{sec-cont} we consider continuous time systems where
oversampled data can lead to small values of $K$ despite an underlying
chaotic dynamics. In Section~\ref{sec-noise} we investigate the issue
of measurement noise.

\begin{rmk} \label{rmk-first}
In the first version of our test, introduced in \cite{GM04}, we defined
$p_c(n)$ and $q_c(n)$ by iterating the extended system
\begin{align*}
p_c(n+1) &= p_c(n) + \phi(n) \cos(\vartheta_c(n)) \nonumber\\
q_c(n+1) &= q_c(n) + \phi(n) \sin(\vartheta_c(n)) \nonumber\\
\vartheta_c(n+1) &= \vartheta_c(n)+c+\alpha \, \phi(n)\; .
\end{align*}
The current version of the test corresponds to the case $\alpha=0$.
As shown in \cite{GM05}, the test with $\alpha=0$ is less sensitive to
measurement noise.
\end{rmk}

\begin{rmk} \label{rmk-valid}
It can be rigorously shown that (i) $p_c(n)$ and $q_c(n)$ are bounded
if the underlying dynamics is regular, i.e.  periodic or quasiperiodic 
and (ii) $p_c(n)$ and $q_c(n)$ behave asymptotically like Brownian
motion for large classes of chaotic dynamical systems.

In \cite{GM04} we used results of~\cite{FMT03,MN04,NMA01} to prove
this for the case $\alpha \neq 0$ in Remark~\ref{rmk-first}. In the
case $\alpha=0$ these results are not applicable; nevertheless in
\cite{GMprep} we cover the case $\alpha=0$ under even weaker 
assumptions on the underlying dynamics.
\end{rmk}

\begin{rmk} \label{rmk-third}
In \cite{GM04,GMprep} it was shown that the results are valid for
almost all observables $\phi$. Of course, the choice of the observable
$\phi$ influences the rate of convergence (but not the limiting value
$K=0$ or $K=1$). From a practical point of view we found that changing
the observable does not greatly alter the computed value of $K$.
\end{rmk}


\section{Computation of the mean square displacement}
\label{sec-MSQ}
For a given time series $\phi(j)$ with $j=1,\ldots ,N$, we compute the
mean square displacement of the translation variables $p_c(n)$ and
$q_c(n)$ defined in~\eqref{eq-p} for several values of
$c\in(0,\pi)$. The mean square displacement is defined as
\begin{align}
M_c(n) = \lim_{N\to\infty}\frac{1}{N}
\sum_{j=1}^{N}[p_c(j+n)-p_c(j)]^2\, + \,
[q_c(j+n)-q_c(j)]^2 \; .
\label{e.MSQ}
\end{align}
Note that this definition requires $n\ll N$. In
\cite{GM05} we calculated the mean square displacement
using directly the definition (\ref{e.MSQ}). The limit is assured by
calculating $M_c(n)$ only for $n\le n_{\rm{cut}}$ where
$n_{\rm{cut}}\ll N$. In practice we find that $n_{\rm{cut}}=N/10$
yields good results.

The test for chaos is based on the growth rate of $M_c(n)$ as a
function of $n$.  In the following, we use analytical expressions
derived in \cite{GMprep} to formulate a modified mean square
displacement $D_c(n)$ which exhibits the same asymptotic growth as
$M_c(n)$ but with better convergence properties.

Under mild assumptions on the underlying dynamical
system, described in Remark~\ref{rmk-rho} below, for each $c\in(0,\pi)$,
\begin{align}
\label{e-Mcn}
M_c(n) = V\!(c)\, n + V_{\rm{osc}}(c,n) + e(c,n)\;,
\end{align}
where $e(c,n)/n \to0$ as $n\to\infty$ uniformly in $c\in(0,\pi)$ and
\begin{align*}
V_{\rm{osc}}(c,n) = (E\phi)^2 \frac{1-\cos nc}{1-\cos c}\;.
\end{align*}
The expectation $E\phi$ is given by
\begin{align*}
E\phi=\lim_{N\to\infty}\frac{1}{N}\sum_{j=1}^N \phi(j)\; .
\end{align*}

The form (\ref{e-Mcn}) suggests an improvement for the test: We can
subtract the explicit term $V_{\rm{osc}}(c,n)$ from the mean square
displacement and introduce
\begin{align}
D_c(n) = M_c(n)-V_{\rm{osc}}(c,n)\; .
\label{e.MSQNEW}
\end{align}
Note that the asymptotic growth rates of $M_c(n)$ and $D_c(n)$ are the
same.\\

\begin{figure}[htb]
\centerline{%
\includegraphics[width=.495\textwidth,angle=0]{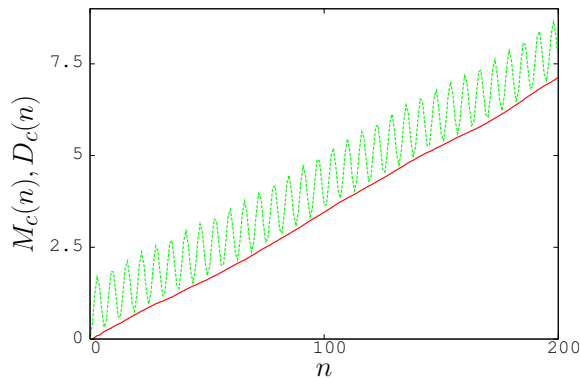}
}
\caption{
Plot of mean square displacement versus $n$ for the logistic map with
$\mu=3.91$ corresponding to chaotic dynamics. The oscillating (green)
curve is the original mean square displacement $M_c(n)$ as defined in
(\ref{e.MSQ}); the straighter (red) curve is the modified mean square
displacement $D_c(n)$ as defined in (\ref{e.MSQNEW}). We used $2000$
data points and computed $M_c(n)$ and $D_c(n)$ for $n=1,\ldots,200$
and $c=1.0$.}
\label{fig-MSQ}
\end{figure}

In Fig.~\ref{fig-MSQ} we show the two mean square displacements
$M_c(n)$ and $D_c(n)$ for the logistic map $x_{n+1}=\mu x_n(1-x_n)$
with $\mu=3.91$ (which corresponds to chaotic dynamics) and an
arbitrary value of $c=1.0$. Evidently, the subtraction
of the oscillatory term $V_{\rm{osc}}(c,n)$ regularizes the linear
behaviour of $M_c(n)$. This allows a much better determination of the
asymptotic growth rate $K_c$.

\begin{rmk} \label{rmk-rho}
The autocorrelation function for the observation $\phi(j)$ is given by 
\[
\rho(k)=E(\phi(1)\phi(k+1))-(E\phi)^2, \enspace k=0,1,2\ldots
\]
Provided the autocorrelations are absolutely summable (that is,
$\sum_{k=0}^\infty |\rho(k)|<\infty$) then equation (\ref{e-Mcn}) is
valid, and moreover, the error term $e(c,n)$ decays uniformly in
$c\in(0,\pi)$ (see for example~\cite{GMprep}).  It is for this reason
that the test based on $D_c(n)$ greatly outperforms the test based on
$M_c(n)$.

Furthermore, the absolute summability condition guarantees~\cite{GMprep} 
that
\begin{align} 
\label{eq-S}
V(c)=\sum_{k=-\infty}^\infty e^{ikc}\rho(|k|)
=\lim_{n\to\infty}\frac1n E \Bigl|\sum_{j=0}^{n-1}e^{ijc} \phi(j) \Bigr|^2
\end{align}
for all $c\in(0,2\pi)$.  This result follows from the Birkhoff ergodic
theorem, the Wiener-Khintchine theorem, and standard calculations.  In
particular, the slope $V(c)$ of the mean square displacement is
identified with the power spectrum.

For nonmixing systems, the error term $e(c,n)$ no longer decays to
zero and there are further oscillatory terms in addition to
$V_{\rm{osc}}(c,n)$.  Nevertheless, the identification~\eqref{eq-S}
remains valid for nonmixing systems under very weak conditions~\cite{MGapp}.

More importantly from the point of view of the test for chaos, working
with $D_c(n)$ remains highly advantageous even for nonmixing systems.
This is illustrated for the logistic map in Fig.~\ref{fig-TEST} later
in this paper,
\end{rmk}

\section{Computation of $K_c$}
\label{sec-Kc}

Having calculated the modified mean square displacement $D_c(n)$ for
$n=1,2,\ldots,n_{\rm{cut}}$, the next step is to estimate the
asymptotic growth rate $K_c$.  We have tried out two different
methods: a {\em regression} method and a {\em correlation} method,
described in subsections~\ref{sec-regress} and~\ref{sec-corr} below.


\subsection{Regression method}
\label{sec-regress}

The regression method consists of linear regression for the log-log
plot of the mean square displacement.  In \cite{GM05} we used the
original mean square displacement $M_c(n)$, so the asymptotic growth
rate $K_c$ is given by the definition
\begin{align*}
K_c=\lim_{n\to\infty}\frac{\log {M}_c(n)}{\log n} \; .
\end{align*}
Numerically, $K_c$ is determined by fitting a straight line to the
graph of $\log M_c(n)$ versus $\log n$ through minimizing the absolute
deviation \cite{NRinC}.

In Section~\ref{sec-MSQ}, we demonstrated the superiority of the
modified mean square displacement $D_c(n)$ when compared to $M_c(n)$,
so it is natural to apply the regression method to $D_c(n)$.  Whereas
$M_c(n)$ is strictly positive, $D_c(n)$ may be negative due to the
subtraction of the oscillatory term $V_{\rm{osc}}(c,n)$. Hence, we set
\begin{align*}
\tilde D_c(n)=D_c(n)-\min_{n=1,\dots,n_{\rm cut}}D_c(n)\;,
\end{align*}
and obtain the asymptotic growth rate
\begin{align*}
K_c=\lim_{n\to\infty}\frac{\log \tilde D_c(n)}{\log n}\; .
\end{align*}
Again, $K_c$ can be determined numerically by regression (minimizing
the absolute deviation) for the graph of $\log \tilde D_c(n)$ versus
$\log n$.

\begin{figure}[htb]
\centerline{
\includegraphics[width=.495\textwidth]{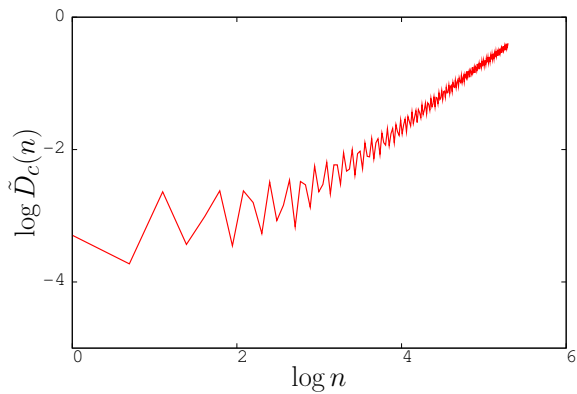}
}
\caption{
Plot of $\log \tilde D_c(n)$ as a function of $\log n$ for the
logistic map at $\mu=3.62$. We used $N=2000$ and calculated the mean
square displacement up to $n_{\rm{cut}}=N/10$.}
\label{fig-logM}
\end{figure}

\begin{rmk}  
\label{rmk-regress} 
Minimizing the absolute deviation is preferable when compared to the
usual least square method as the latter assigns a higher weight to
outliers. Since the linear behaviour of the mean square displacement
is only given asymptotically, one typically encounters outliers for
small values of $n$. We find that it is usually sufficient to use the
absolute deviation for estimating $K_c$, and that it is not necessary
to employ more complicated higher-order regression methods such as the
method by Yohai
\cite{Yohai}.

The finite value of the $o(n)$-term $e(c,n)$ in the definition of
$M_c(n)$ in (\ref{e-Mcn}) leads to a distortion for small values of
$n$.  In such situations, one typically observes a flattening of the
slope of $\log {M}_c(n)$ or $\log \tilde D_c(n)$ as illustrated in
Fig.~\ref{fig-logM}. It is those values for small $n$ of $\log
{M}_c(n)$ (or $\log \tilde D_c(n)$) which would be overestimated in a
least square fit.
\end{rmk}


\subsection{Correlation method}
\label{sec-corr}

We now present an alternative method for determining $K_c$ from the
mean square displacement.  (The method is described in terms of
$D_c(n)$, but we could use $M_c(n)$ instead.)

Form the vectors $\xi=(1,2,\dots,n_{\rm cut})$ and
$\Delta=(D_c(1),D_c(2),\dots,D_c(n_{\rm cut}))$.  Given vectors $x$,
$y$ of length $q$, we define covariance and variance in the usual way:
\begin{align*}
& {\rm cov}(x,y)   = \frac1q\sum_{j=1}^q (x(j)-\bar x)(y(j)-\bar y), 
\quad \text{where} \enspace
\bar x = \frac1q\sum_{j=1}^q x(j)\; , \\
& {\rm var}(x)  ={\rm cov}(x,x)\; .
\end{align*}
Now define the correlation coefficient
\begin{align*}
K_c = {\rm corr}(\xi,\Delta)=
\frac{{\rm{cov}}(\xi,\Delta)}{\sqrt{{\var(\xi)}{\var(\Delta)}}}\in[-1,1]\;.
\end{align*}
This quantity measures the strength of the correlation of $D_c(n)$
with linear growth.  Again, it can be shown rigorously~\cite{GMprep}
that under weak conditions on the underlying dynamics (as described in
Remark~\ref{rmk-valid}) we obtain $K_c=0$ for regular dynamics and
$K_c=1$ for chaotic dynamics.

\begin{figure}[htb]
\centerline{%
\includegraphics[width=.495\textwidth,angle=0]{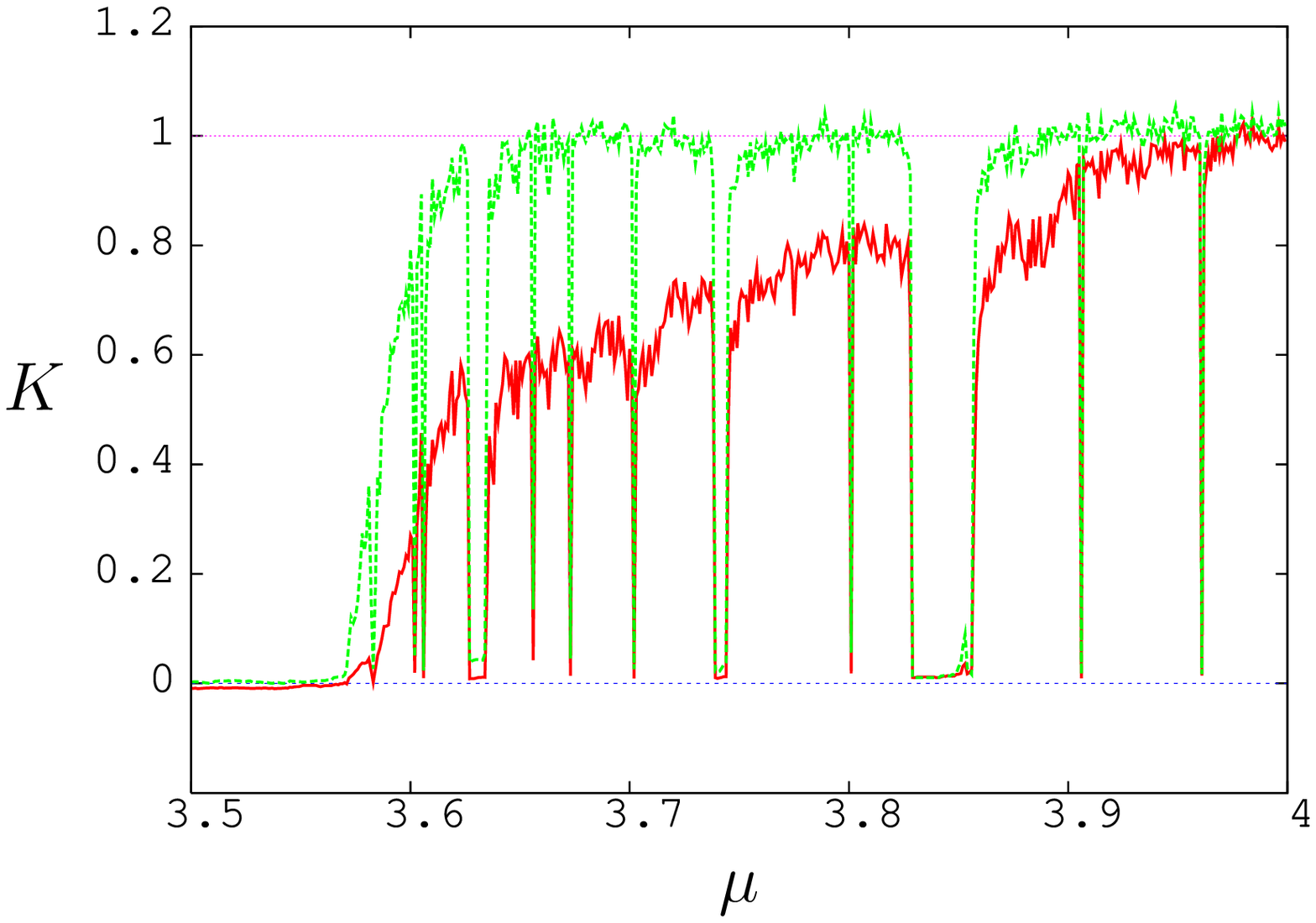}
\includegraphics[width=.495\textwidth,angle=0]{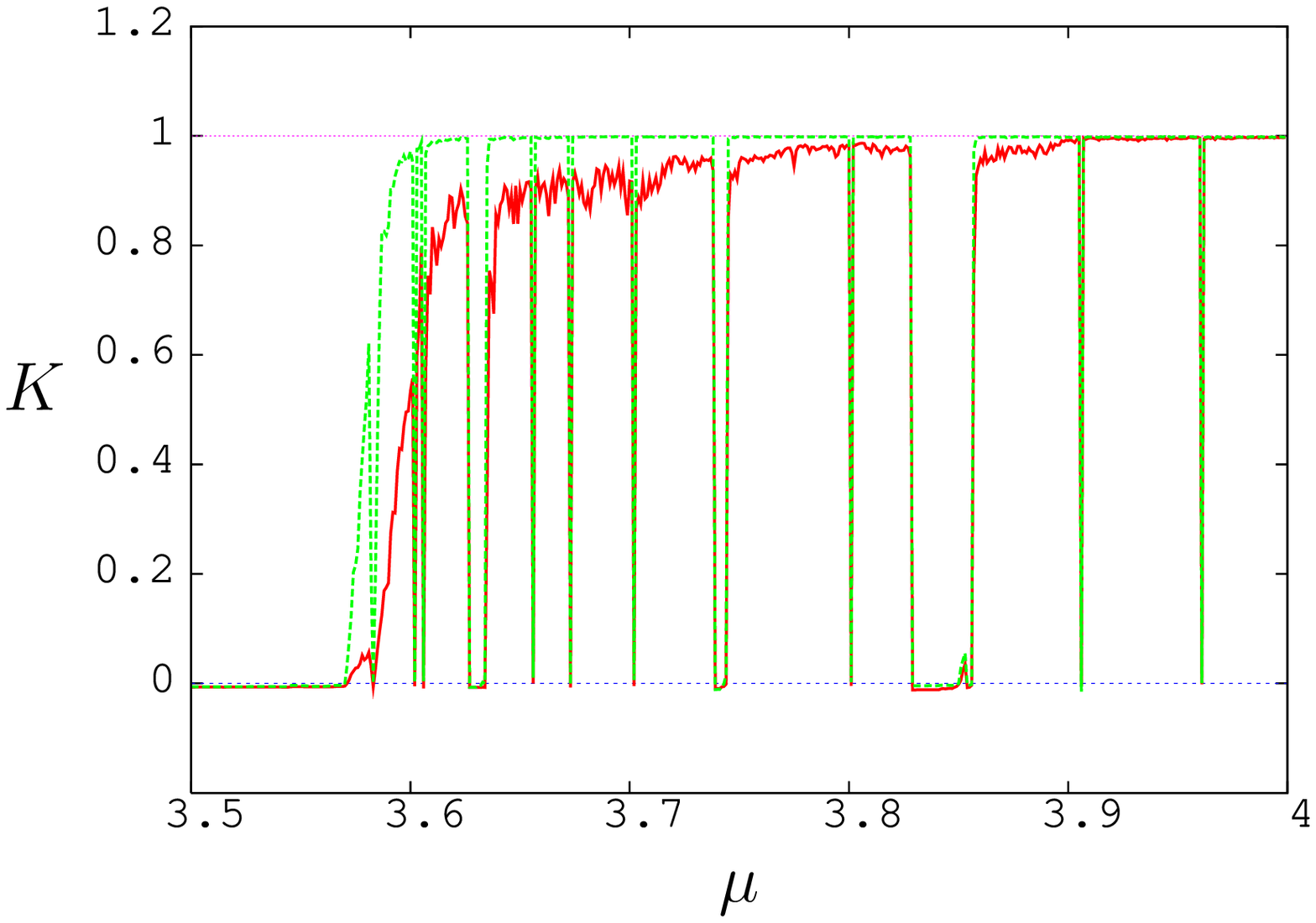}
}
\caption{
Plot of $K$ versus $\mu$ for the logistic map with $3.5\le \mu\le 4$
increased in increments of $0.001$. We used $2000$ data points. The
darker (red) lines are obtained by using the original definition of
the mean square displacement $M_c(n)$ in (\ref{e.MSQ}). The lighter
(green) lines are obtained by using the modified mean square
displacement $D_c(n)$ in (\ref{e.MSQNEW}).  The resulting values of
$K$ are shown for the regression method (left) and the correlation
method (right).  The horizontal lines (blue and magenta) indicate the
cases $K=0$ and $K=1$. We used $N_c=100$ values of $c$.  }
\label{fig-TEST}
\end{figure}

In practical terms, the correlation method greatly outperforms the
regression method.  This is evident from Fig.~\ref{fig-TEST} which
compares the regression method and the correlation method (using both
$M_c(n)$ and $D_c(n)$) for the logistic map.


\section{Choice of $c$ and determination of $K$}
\label{sec-c}

\begin{figure}[htb]
\centerline{%
\includegraphics[width=.33\textwidth]{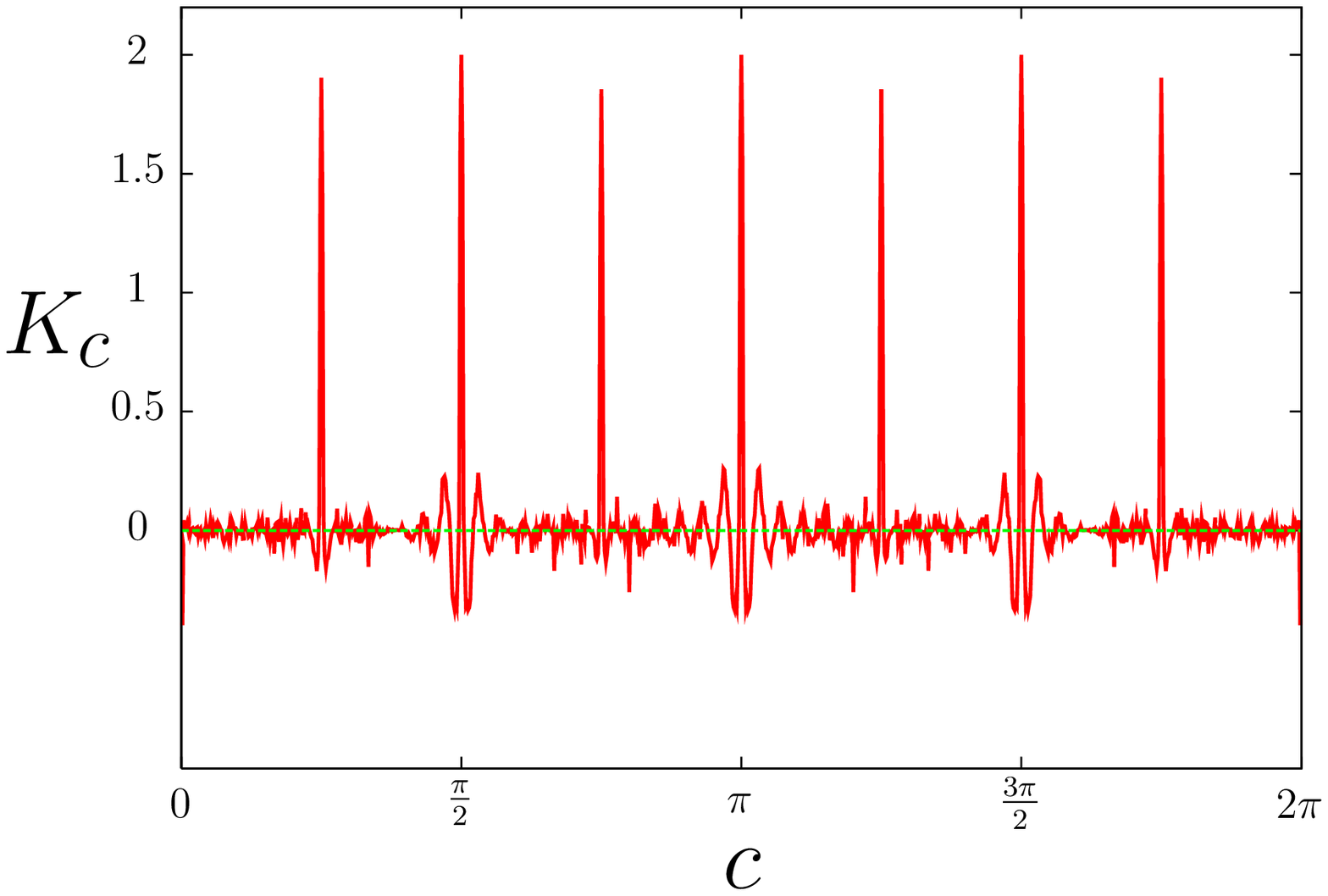}
\includegraphics[width=.33\textwidth]{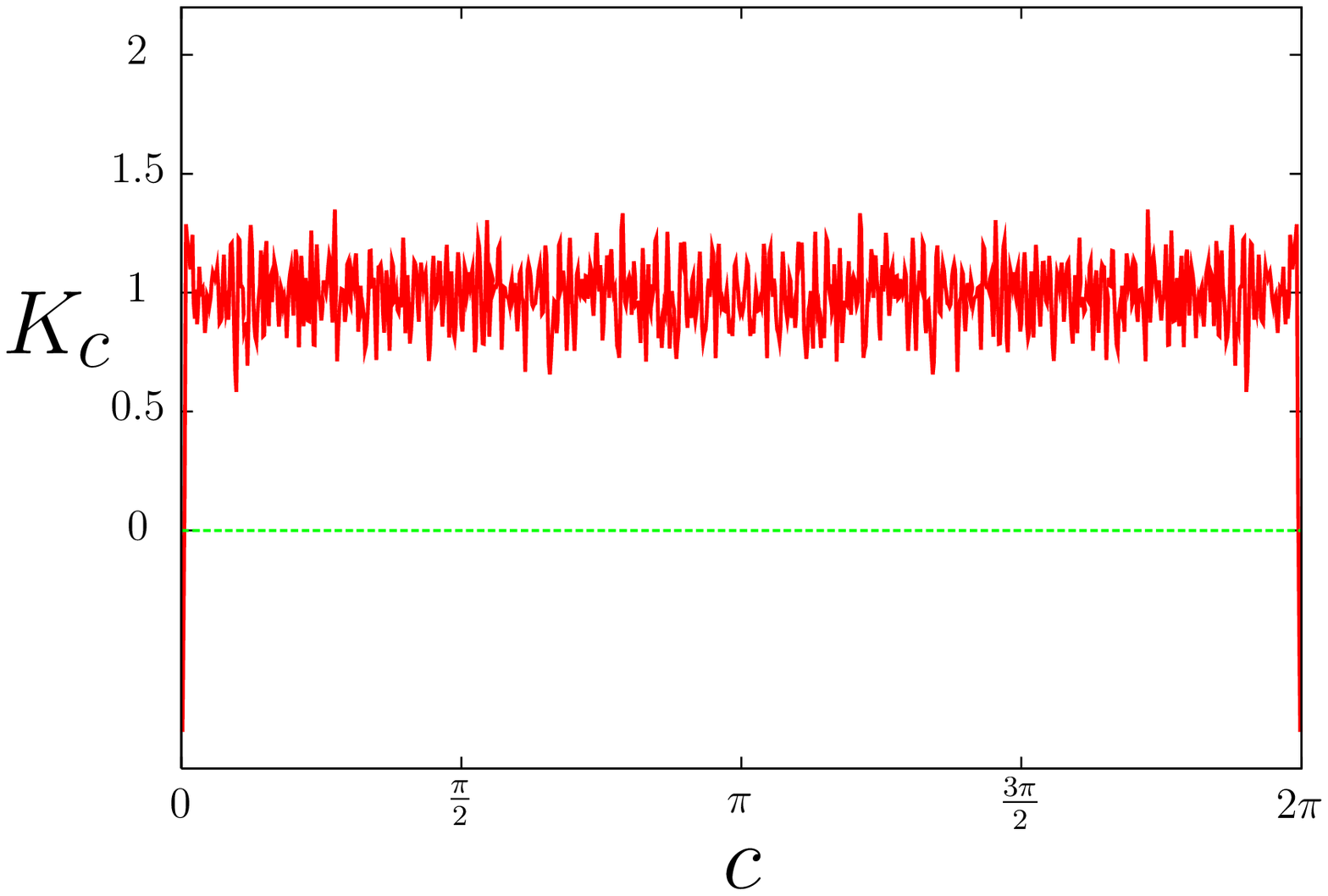}
\includegraphics[width=.33\textwidth]{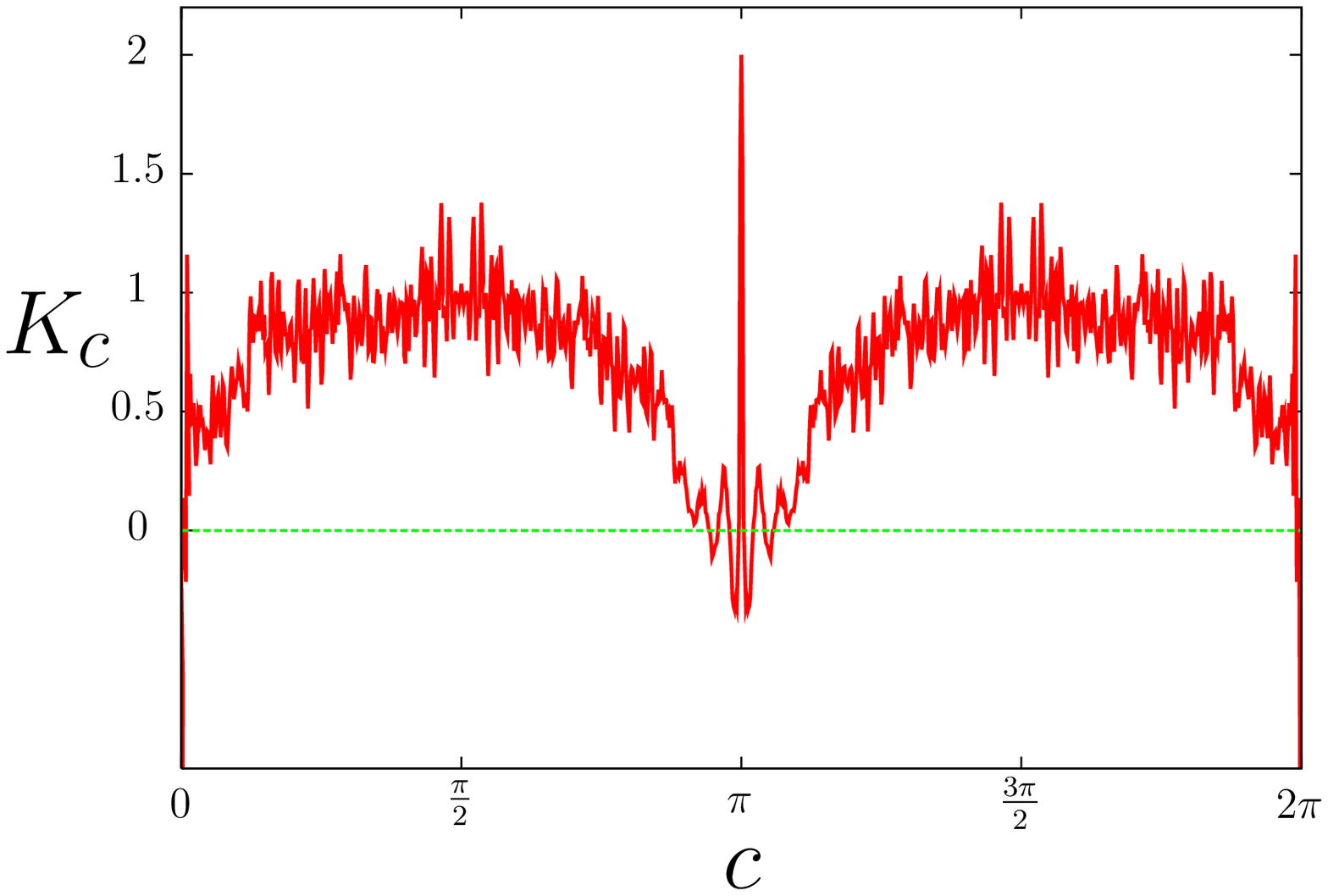}
}

\vspace*{1ex}
\centerline{%
\includegraphics[width=.33\textwidth]{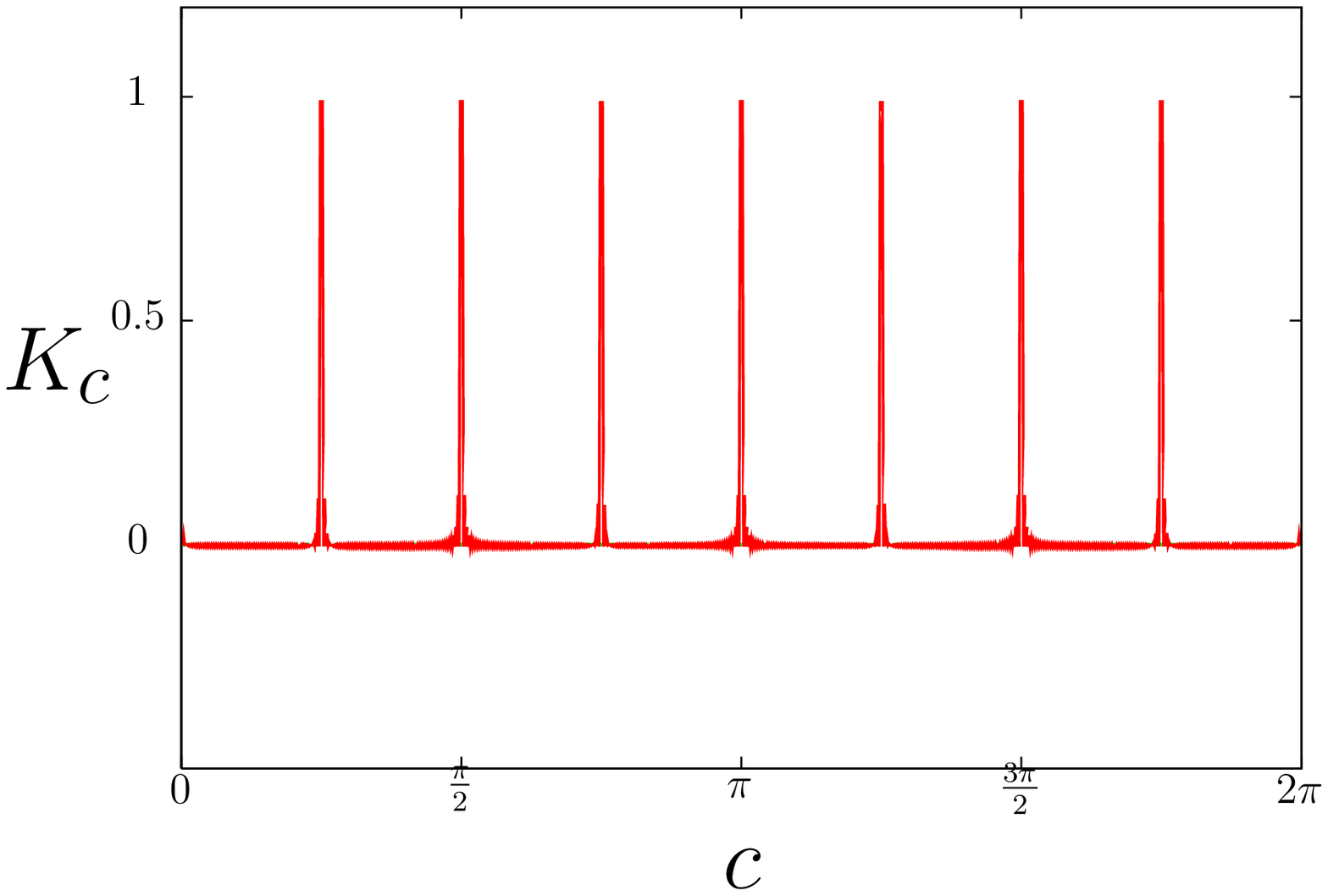}
\includegraphics[width=.33\textwidth]{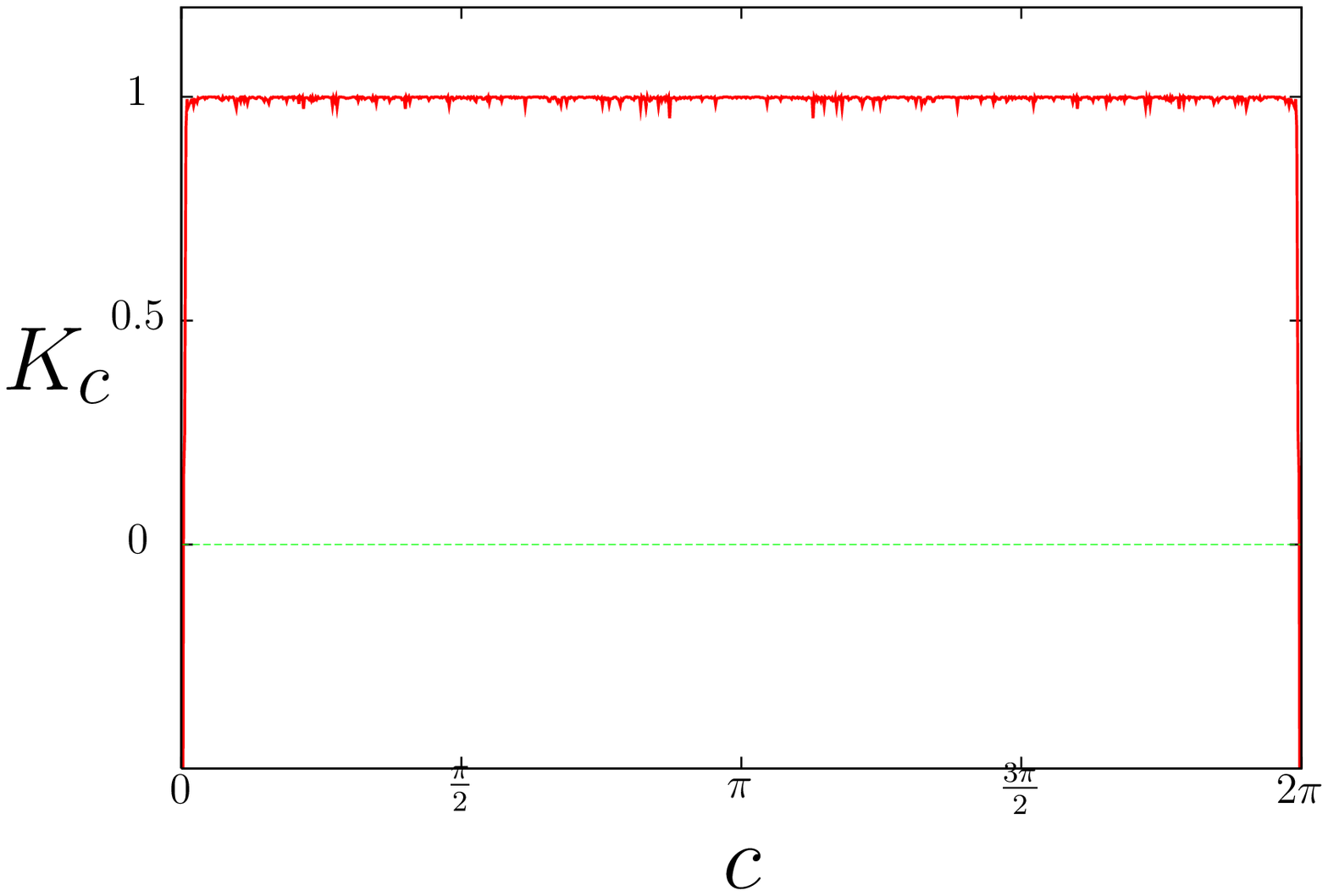}
\includegraphics[width=.33\textwidth]{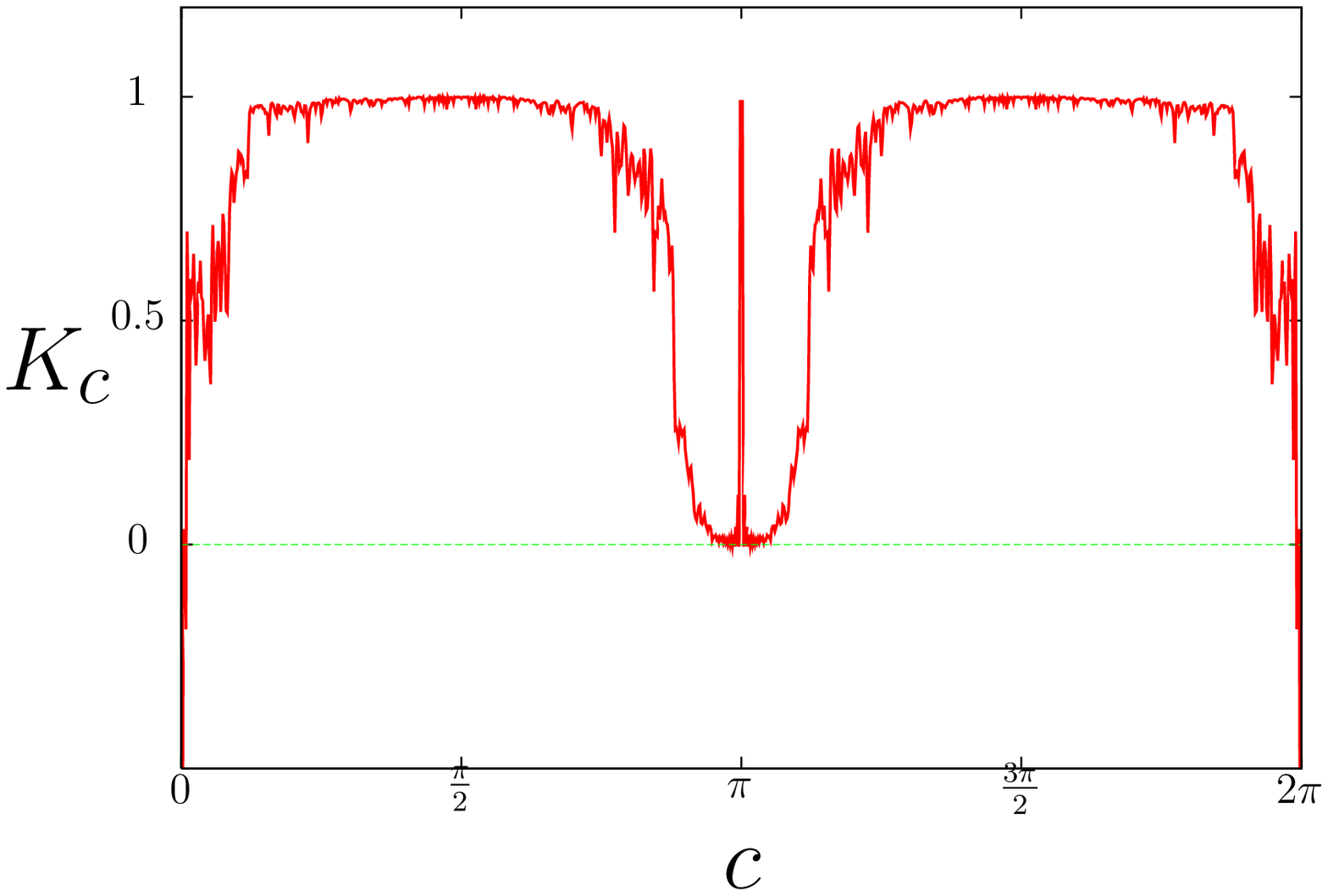}
}
\caption{
Plot of $K_c$ versus $c$ for the logistic map calculated using the
regression method (top) and correlation method (bottom). We used
here $N=5000$ data points, and $1000$ equally spaced values for $c$.
Left: $\mu=3.55$ corresponding to regular
dynamics; Middle: $\mu=3.9$ corresponding to chaotic dynamics; Right:
$\mu=3.6$ corresponding to chaotic but non-mixing dynamics.}
\label{fig-cscan}
\end{figure}

In Fig.~\ref{fig-cscan} we show the asymptotic growth rate $K_c$ as a
function of $c$ for regular and chaotic dynamics. In the case of
periodic dynamics, most values of $c$ yield $K_c=0$ as expected, but
there are isolated values of $c$ for which $K_c$ is large.  (For the
regression method, $K_c\approx2$ at these resonant points.)  These
resonances are easily explained as follows: Equation (\ref{eq-p})
shows that if the Fourier decomposition of the observation $\phi$
contains a term proportional to $\exp (-i \omega k)$, then there is a
resonance at $c=\omega$ where $p_c(n)\sim n$, and hence $M_c(n) \sim
n^2$, irrespective of whether the dynamics is regular or chaotic. For
the plots in Fig.~\ref{fig-cscan}, we have calculated the asymptotic
growth rate using both the regression method described in
Section~\ref{sec-regress} and the correlation method described in
Section~\ref{sec-corr}.

The occurrence of resonances for isolated values of $c$ suggests using
the median of the computed values of $K_c$.  (We use the median rather
than the mean, since the median is robust against outliers associated
with resonances.)

In Fig.~\ref{fig-cscan}c, $K_c$ is shown as a function of $c$ for
$\mu=3.6$ where the dynamics is chaotic but not mixing on the whole
interval $[0,1]$. The actual dynamics in the logistic map oscillates
between two disjoint sets, each of which is mixing, and there is a
resonance at $c=\pi$. At resonance, $p_c(n)\sim n$ and $M_c(n)\sim
n^2$ as before.  Close to resonance, the $p$-$q$ plot eventually
behaves like Brownian motion, but in practice one sees only a small
part of this motion and so $K_c\approx 0$.

\begin{rmk} \label{rmk-c}
Naturally, the choices of $c$ are equally spaced in
Fig.~\ref{fig-cscan}, whereas in applying the test (and throughout the
paper with the exception of Fig.~\ref{fig-cscan}) we choose randomly
sampled values of $c$.
\end{rmk}

To avoid that resonances distort the statistics, we further restrict
the range of randomly sampled values for $c$ to $c\in(\pi/5 ,4 \pi/5)$
for all our computations. The resonance at $c=0$ is inherent to our
test, but it may leak through adjacent values of $c$ as seen in
Fig.~\ref{fig-cscan}c. The further restriction to exclude $\pi$ is not
necessary, but we found it helpful.  (A typical route to chaos is the
Feigenbaum route via period doubling. Here, the parameter ranges for
fixed points and period two points are largest.)

\begin{figure}[htb]
\center{$\begin{array}{cc}
\includegraphics[width=.495\textwidth]{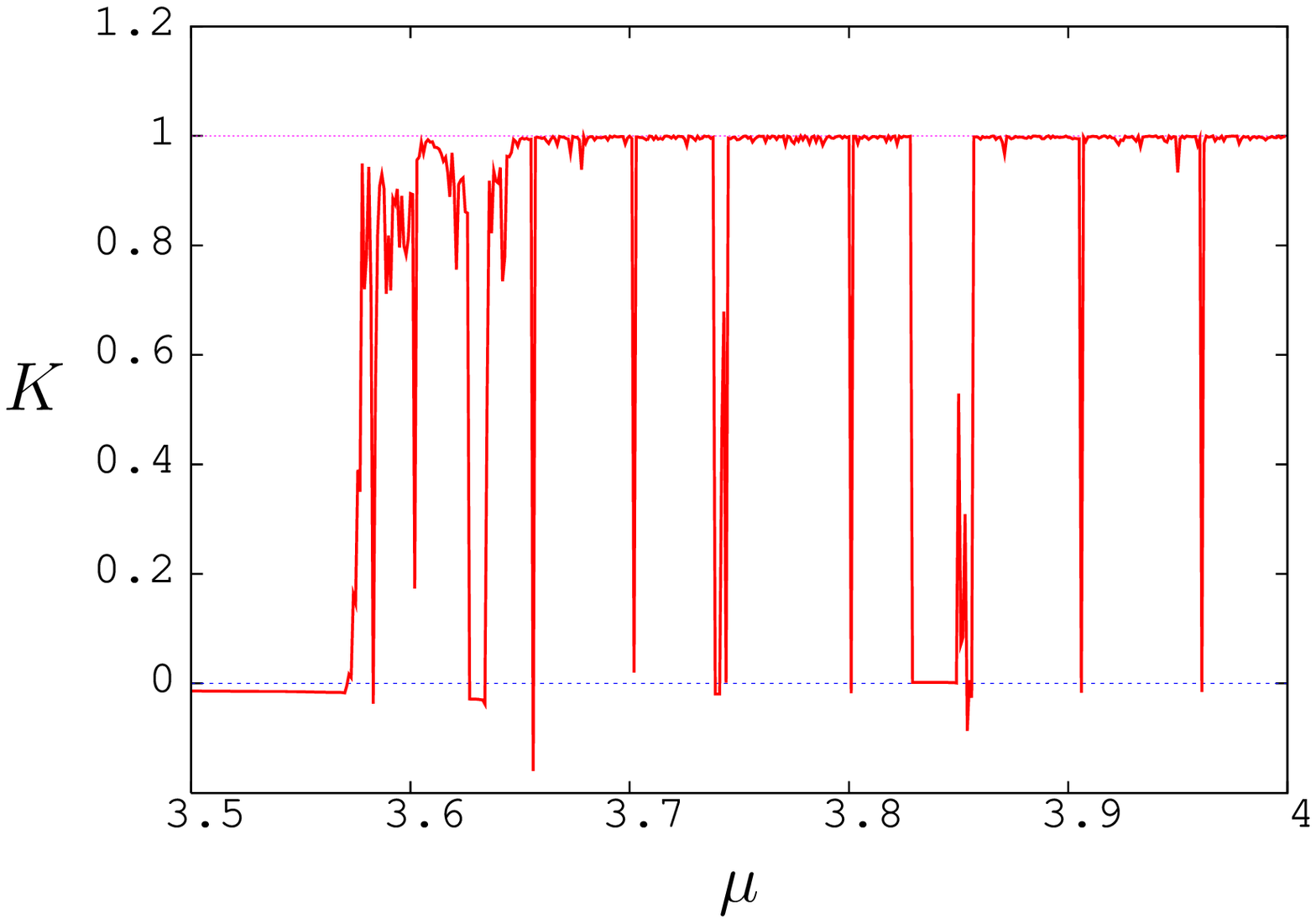}&
\includegraphics[width=.495\textwidth]{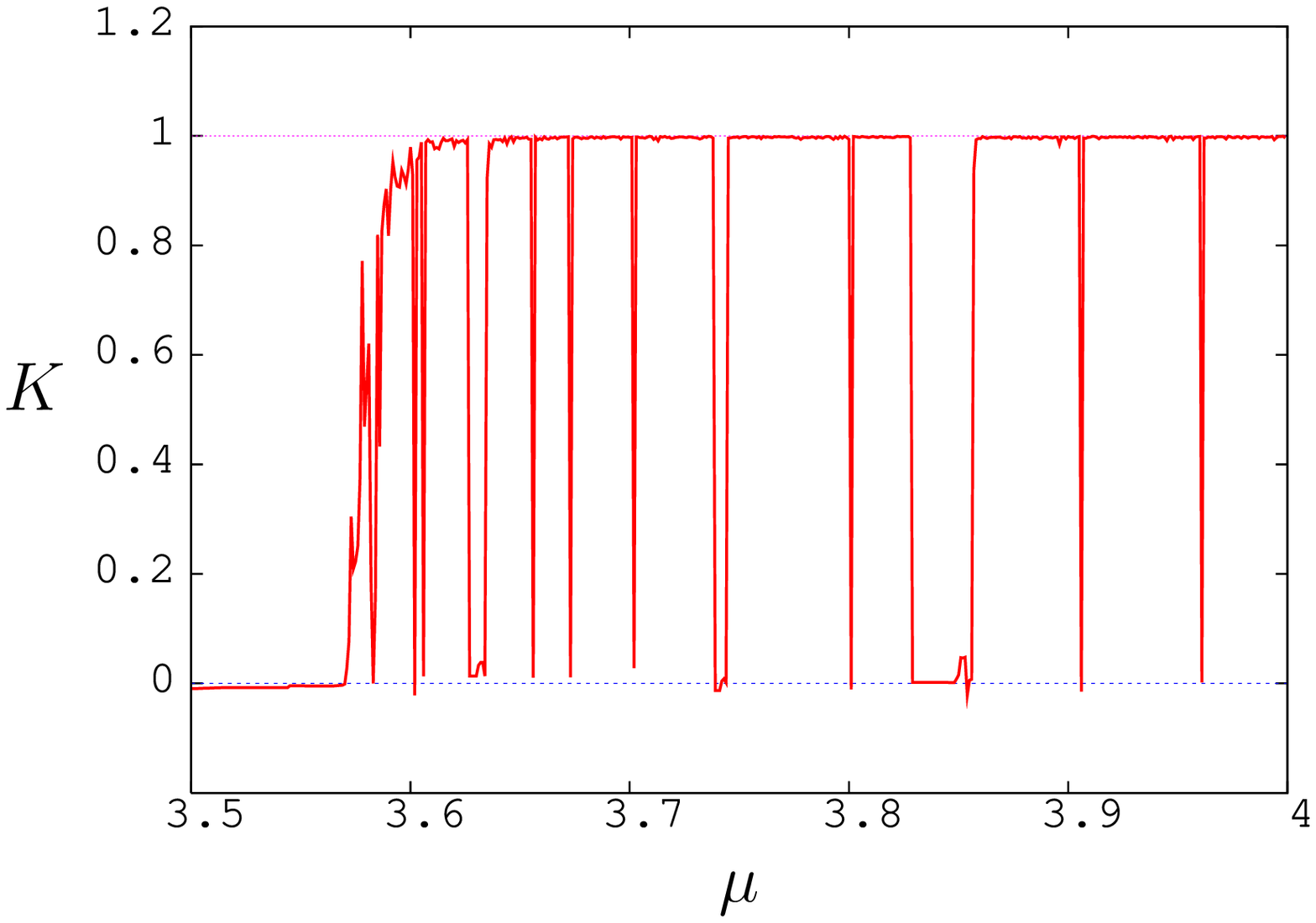}\\
\includegraphics[width=.495\textwidth]{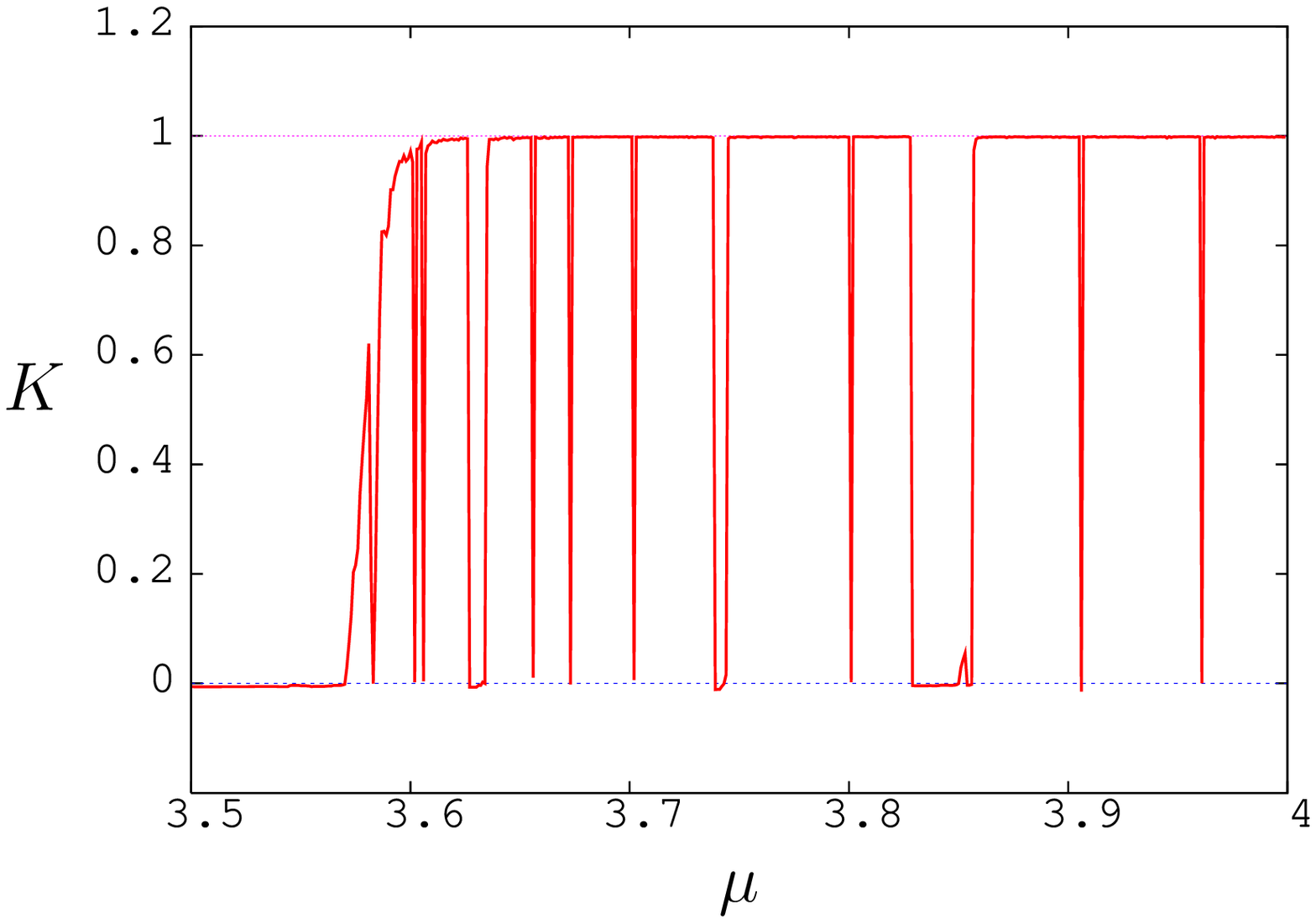}&
\includegraphics[width=.495\textwidth]{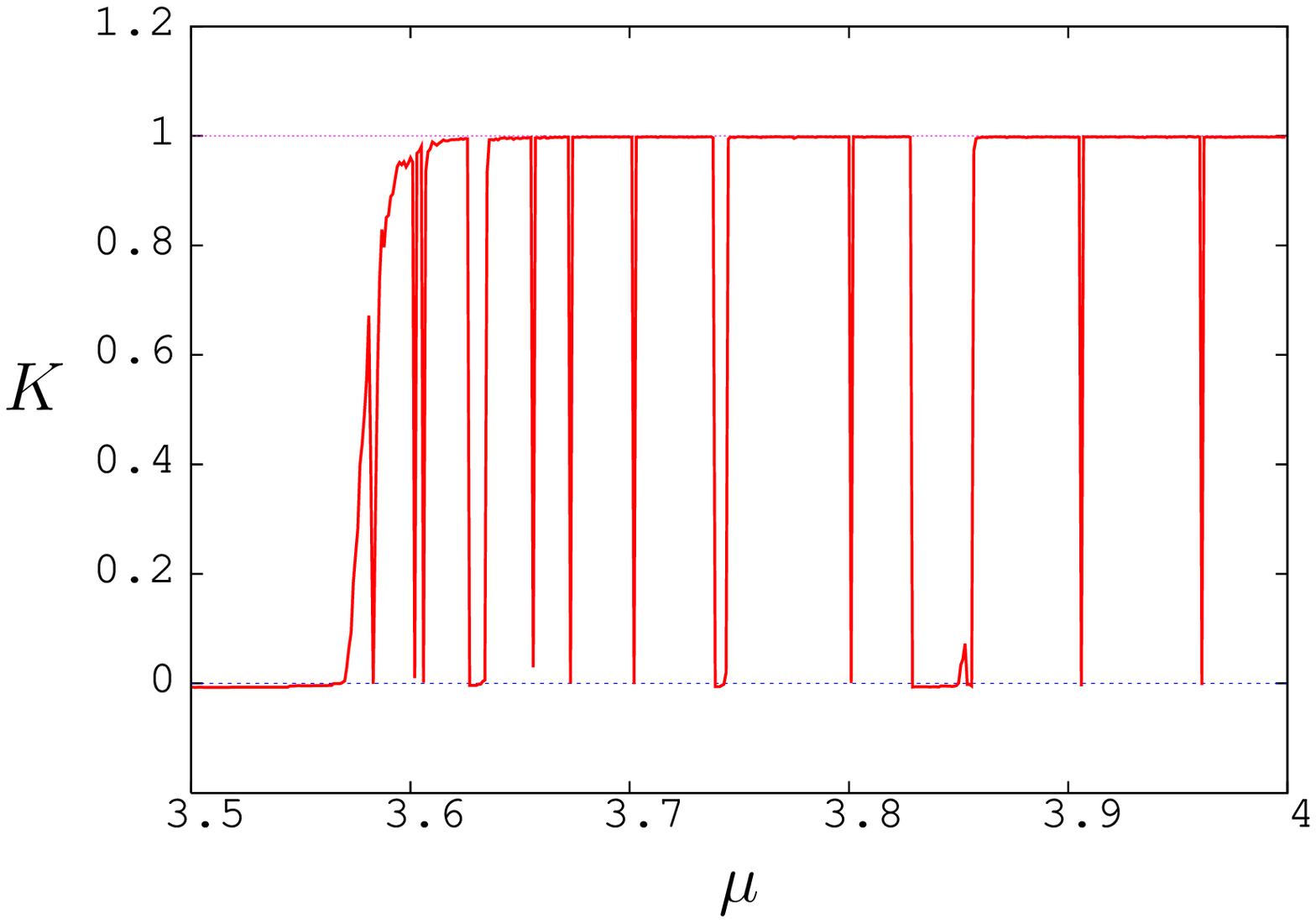}\\
\end{array}$}
\caption{
Plot of $K$ versus $\mu$ for the logistic map using the correlation
method, with $3.5\le \mu\le 4$ increased in increments of $0.001$. We
used $2000$ data points. Upper left: $N_c=1$; Upper right: $N_c=10$;
Lower left: $N_c=100$; Lower right: $N_c=1000$.}
\label{fig-c-Nc}
\end{figure}

In Fig.~\ref{fig-c-Nc}, we show how the result for $K$ depends on the
number $N_c$ of different values of $c$. Here we use the correlation
version of the test to calculate $K_c$ as described in
Section~\ref{sec-corr}. There is no measurable gain in increasing
$N_c$ from $100$ to $1000$ and we find that generally $N_c = 100$
different values of $c$ is sufficient.

\section{Finite size problems}
\label{sec-confidence}

There are three types of finite size effects. First, the time series
needs to be long enough to explore and sample the relevant phase space
area (i.e.\ the attractor).  This is an inherent problem affecting all
tests for chaos. Second, the definition of the mean square
displacement involves a limit which requires $n \ll N$. Accordingly,
we have chosen $n\le n_{\rm{cut}}=N/10$.

Third, the theory developed in \cite{GM04,GMprep} makes statements
about the asymptotic behaviour of $D_c(n)$ (or $M_c(n)$) and as such
requires $n_{\rm cut}$, and hence $N$, to be sufficiently
large. Finite size effect in this context means that for small $n$ the
asymptotic linear growth is not yet dominating, see
Fig.~\ref{fig-logM}. This finite size effect is explored in the
remainder of this section.  From now on, we work exclusively with the
modified mean square displacement $D_c(n)$ and the correlation method.

\begin{figure}[htb]
\centerline{%
\includegraphics[width=.495\textwidth]{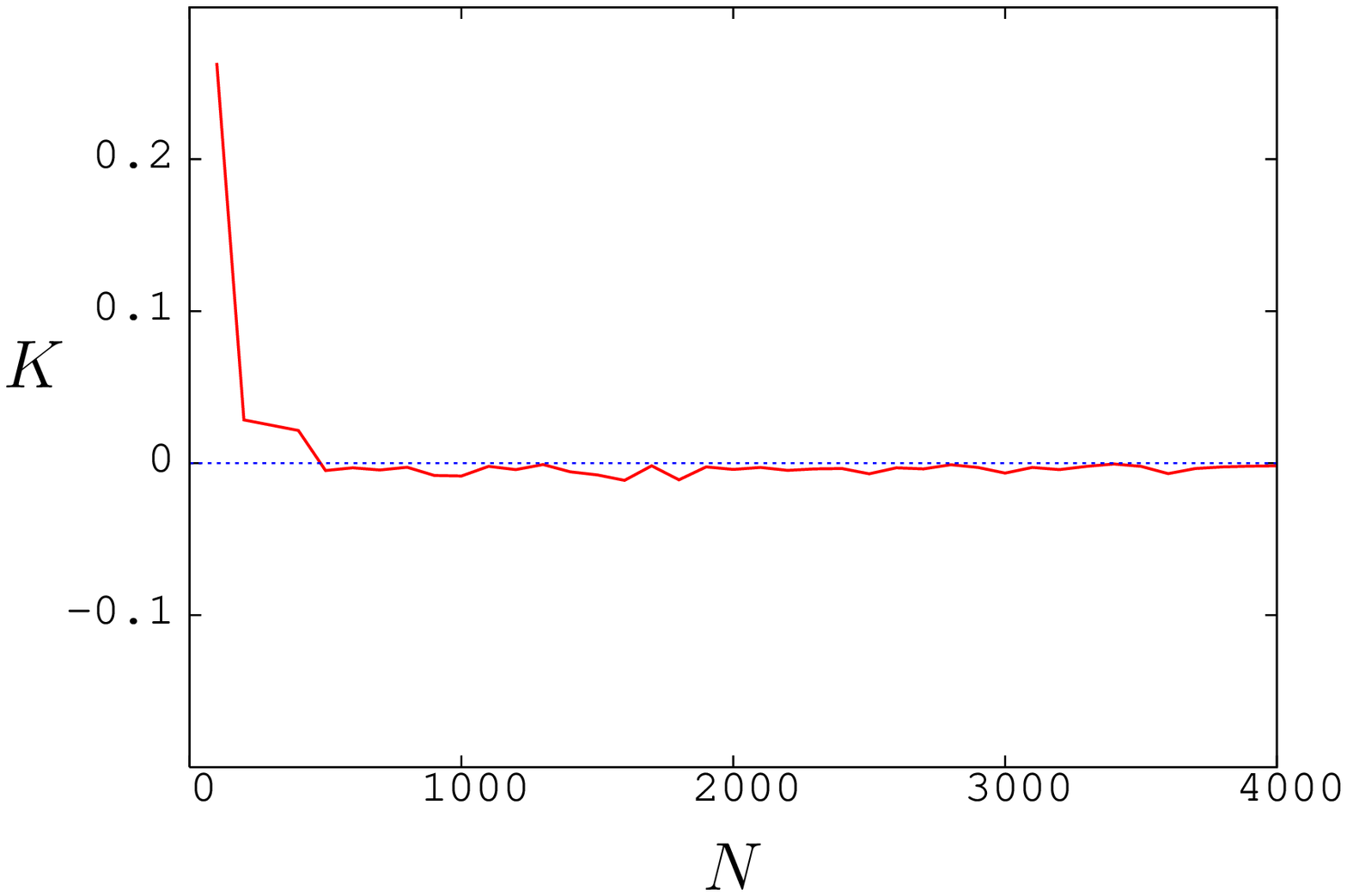}
\includegraphics[width=.495\textwidth]{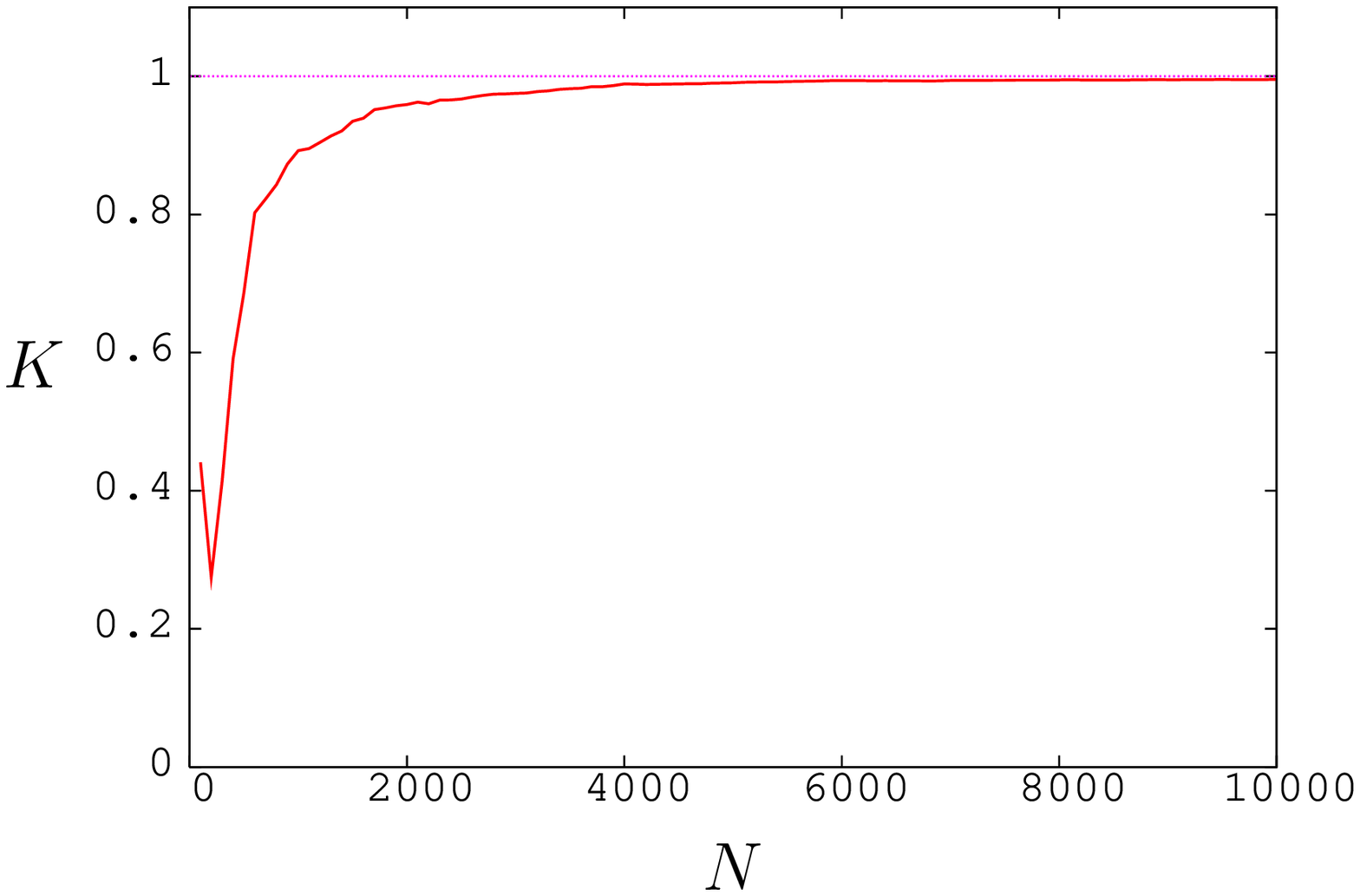}
}
\caption{
Plot of $K$ versus the available amount of data $N$ for the logistic
map. Left: $\mu=3.55$ corresponding to regular dynamics; Right:
$\mu=3.6$ corresponding to chaotic dynamics.}
\label{fig-K_of_N}
\end{figure}

In Fig.~\ref{fig-K_of_N} we show how the value of $K$ depends on the
amount of data used. We can see clearly the convergence towards the
asymptotic values $K=0$ and $K=1$ for regular and chaotic underlying
dynamics, respectively.  (For values of $\mu$ corresponding to
stronger chaotic dynamics well within the chaotic range, the
convergence towards $K=1$ is even more rapid.)

In the case of ``weak chaos'', close to the so called ``edge of
chaos'', longer data sets are required to obtain $K=1$. Weak chaos is
characterized by a slow decay of correlations. This has consequences
for the modified mean square displacement $D_c(n)=V(c) n + o(n)$. For
systems whose auto-correlation function is slowly decaying, it may be
the case that the $o(n)$ term dominates for the available data.  We
illustrate this problem in the context of the logistic map. The
bifurcation parameter $\mu$ takes the value
$\mu=\mu_{\infty}=3.569945672\dots$ at the edge of chaos and for
$\mu=\mu_{\infty}+0.001$ one observes weak chaos.

\begin{figure}[htb]
\centerline{%
\includegraphics[width=.4\textwidth,height=.4\textwidth]{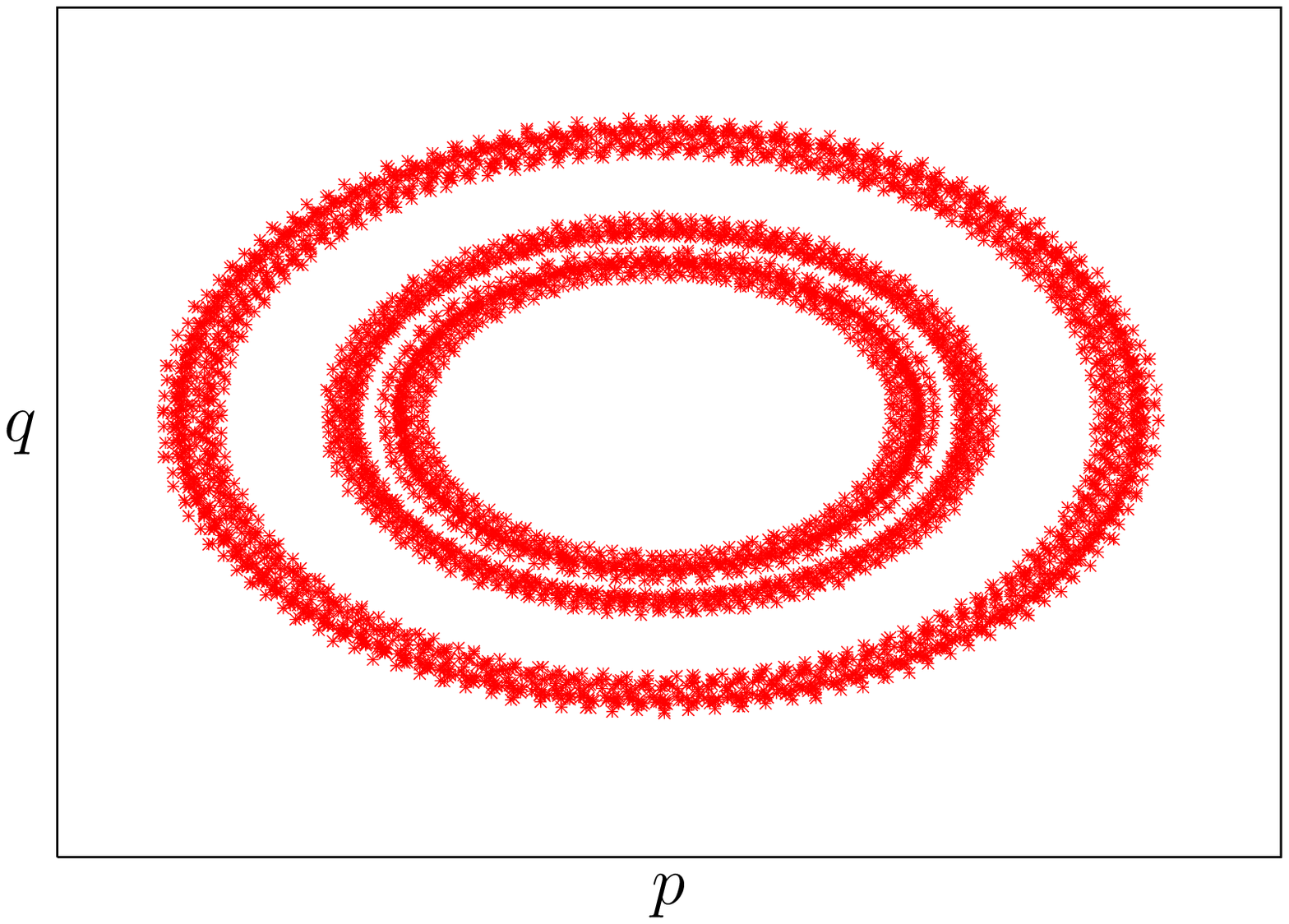}
\qquad
\includegraphics[width=.4\textwidth,height=.4\textwidth]{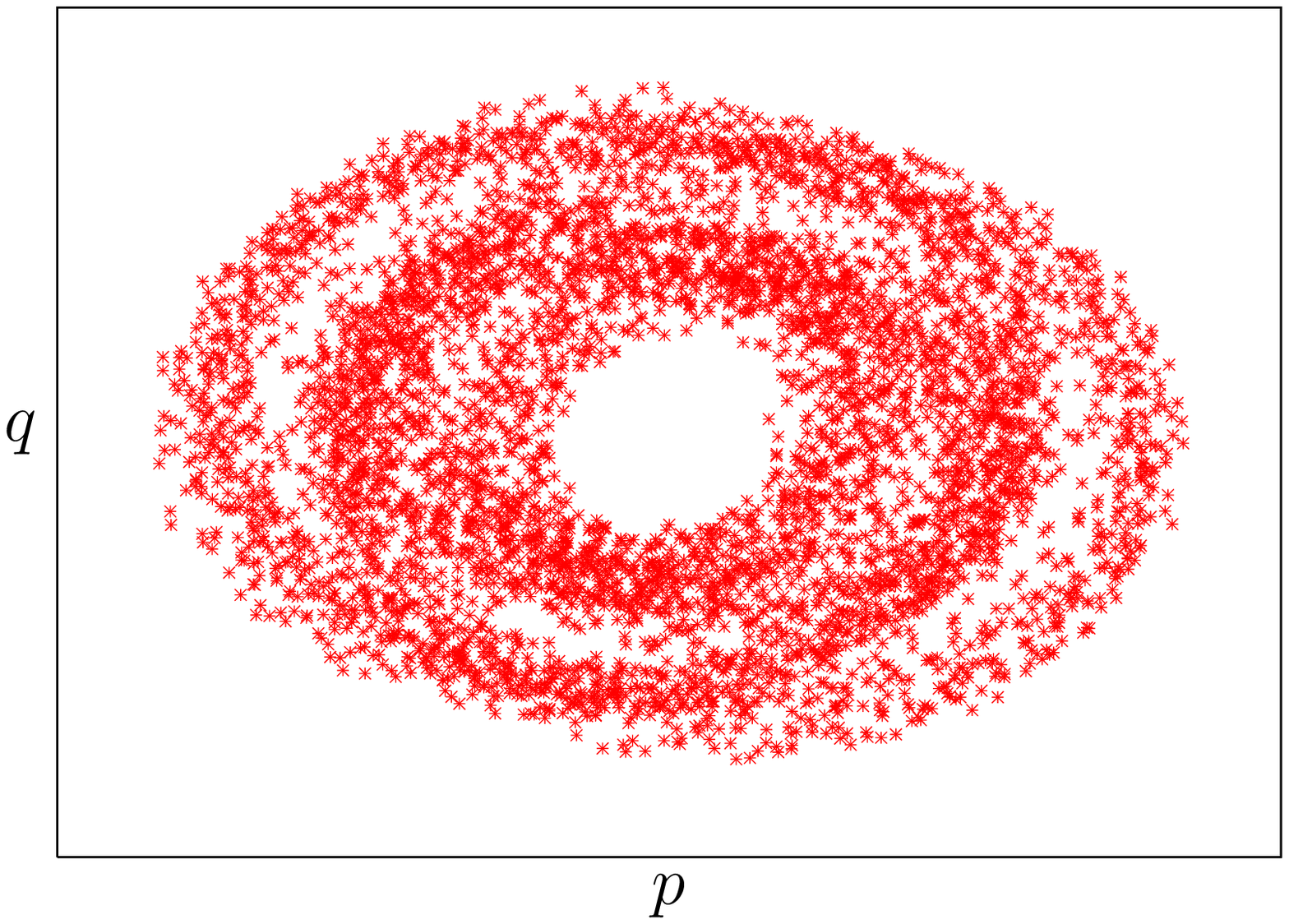}
}
\caption{
Plot of $p$ versus $q$ for the logistic map. Left: $\mu=\mu_{\infty}$;
Right: $\mu=\mu_{\infty}+0.001$. We used $5000$ data points.}
\label{fig-pq2}
\end{figure}

\begin{figure}[htb]
\centerline{%
\includegraphics[width=.475\textwidth]{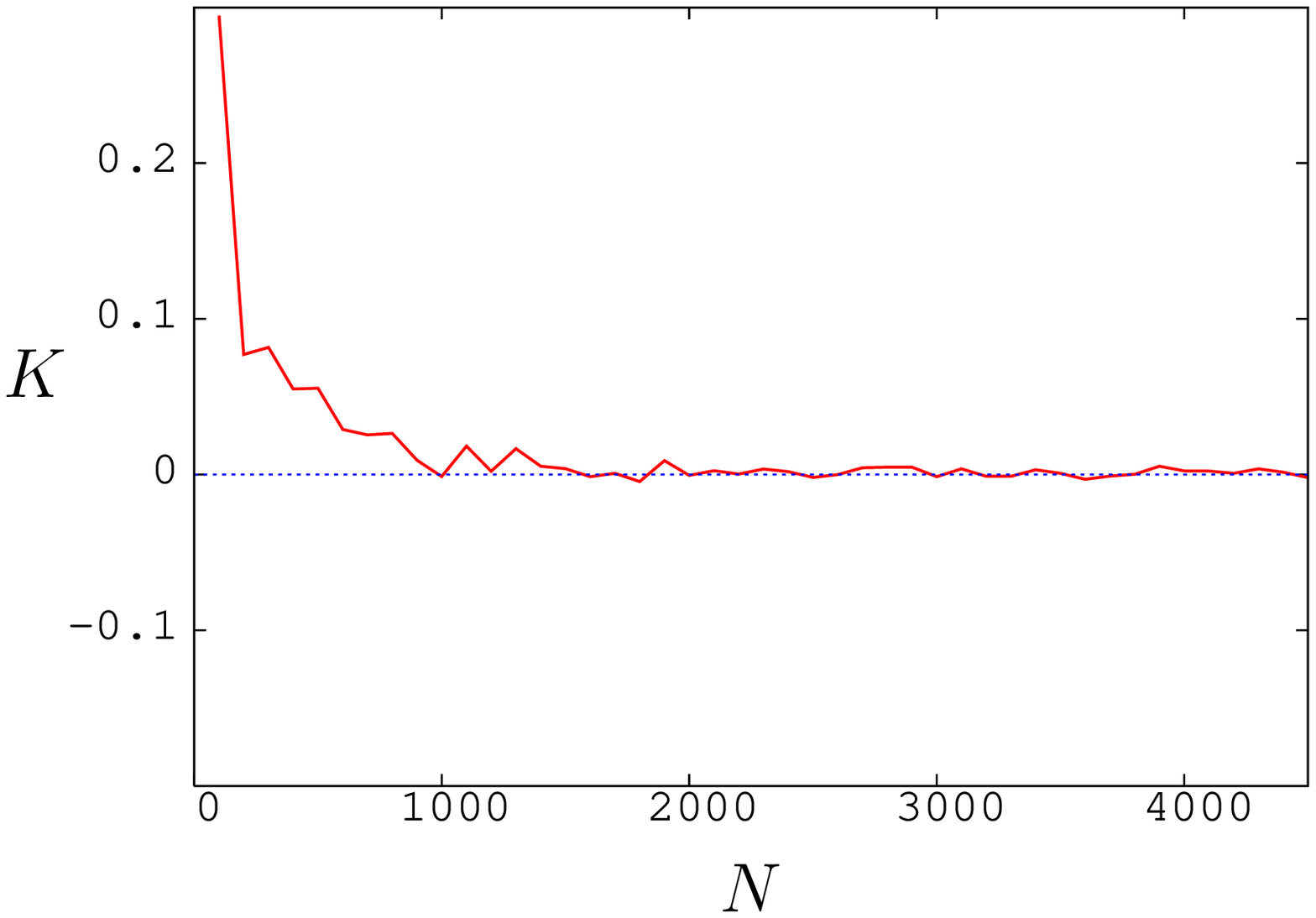}
\includegraphics[width=.495\textwidth]{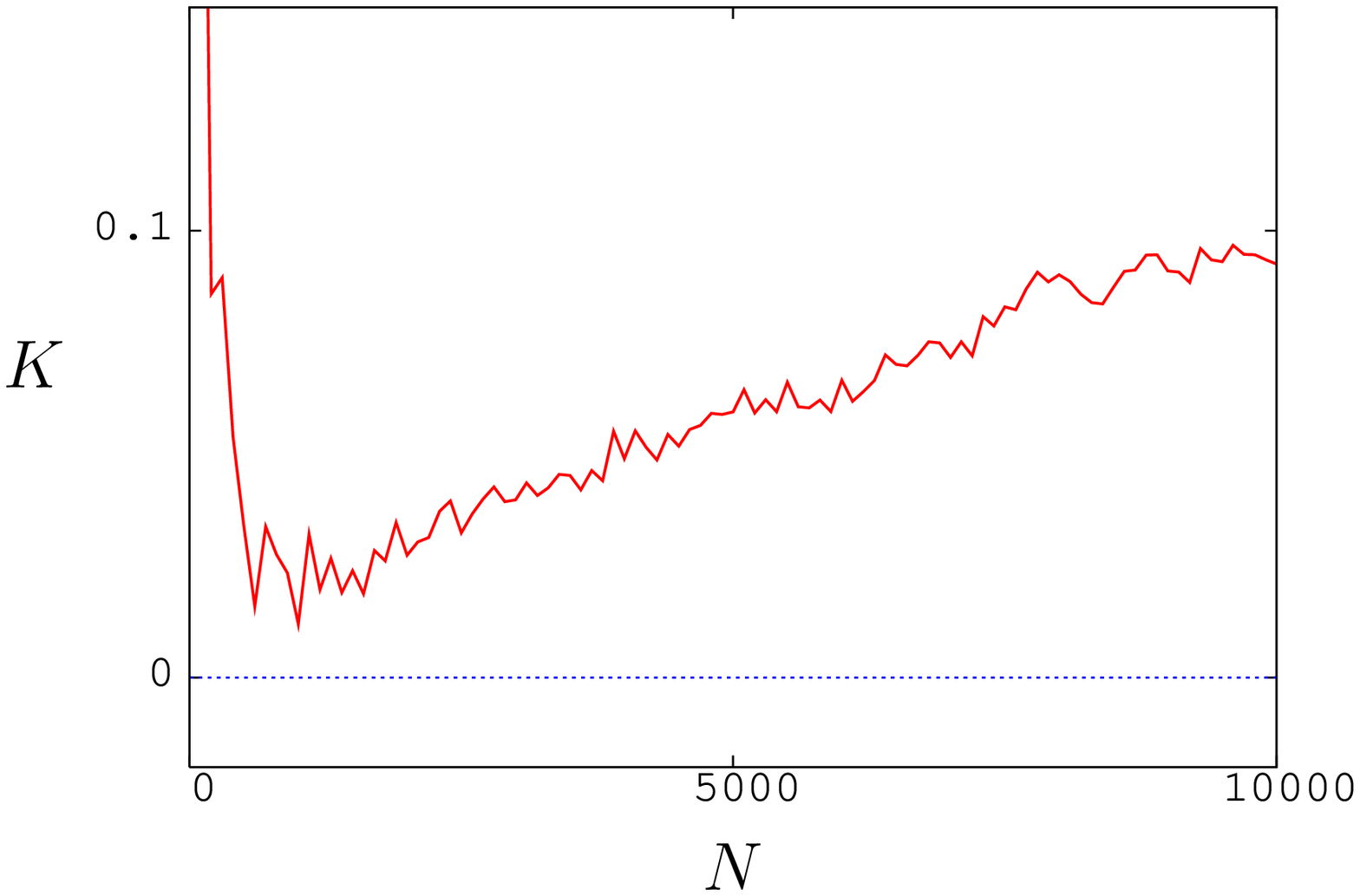}
}
\caption{
Plot of $K$ as a function of $N$ for the logistic map at
$\mu=\mu_\infty$ (left) and $\mu=\mu_\infty+0.001$ (right). Although
the value of $K$ is small in both cases, the behaviour of $K$ as a
function of $N$ distinguishes the two cases.}
\label{fig-Kn}
\end{figure}

It has been erroneously claimed that our test cannot detect weak
chaos, see \cite{HuTung05,GMsub}. In fact, there are two methods
whereby we can distinguish between regular dynamics and weak chaos:
\begin{itemize}
\item[(i)]
By visual inspection of the plot in the $p$-$q$ plane 
generated as in Fig.~\ref{fig-pq2}.
(Note that for longer data sets the dynamics in the $p$-$q$ plane in
Fig.~\ref{fig-pq2}b would look just like Fig.~\ref{fig-pq}b.)
\item[(ii)] By looking at the dependence of $K$ as a function of $N$. 
As illustrated in Fig.~\ref{fig-Kn}, we can 
distinguish weakly chaotic from regular dynamics even when the value
of $K$ is very small -- note that $K=0.027$ for $N=2000$ in the weakly
chaotic case.
\end{itemize}

\clearpage

\section{Continuous time systems}
\label{sec-cont}
              
In the previous sections, the $0$--$1$ test was formulated for discrete
time systems.  For continuous time series $\phi(t)$, there is a
well-known oversampling issue that must be addressed.  In this
section, we discuss this difficulty and how to overcome it.

Given $0<t_1<t_2<t_3<\cdots$ we obtain a discrete time series
$\phi(t_1)$, $\phi(t_2)$, $\phi(t_3),\ldots$ to which the test for
chaos may be applied as in previous sections.  (The sequence $t_j$,
$j\ge1$, should be chosen in a deterministic manner so that the time
series $\phi(t_j)$ is deterministic.)  One method of choosing the
$t_j$ is as the intersection times with a cross-section, so the time
series $\phi(t_j)$ corresponds to observing a Poincar\'e map.  In this
situation, there are no issues with oversampling.

A second, perhaps more usual, approach is to take $t_j=j\tau_s$ where
$\tau_s>0$ is the sampling time.  The time series
$\phi(t_j)=\phi(j\tau_s)$ corresponds to observing the
``time-$\tau_s$'' map associated with the underlying continuous time
system.  If $\tau_s$ is too small, then the system is {\em
oversampled} and this often leads to incorrect results.  To illustrate
the issue of oversampling we study the $3$-dimensional Lorenz system
\begin{align}
\nonumber
\dot x &= {\textstyle10}(y-x)\\
\dot y &= {\textstyle30}\,x-y-xz
\label{lorenz}
\\
\dot z &= xy - {\textstyle\frac83} z \; ,
\nonumber
\end{align}
which exhibits robust chaos.  We have integrated this system with a
time step of $\Delta t=0.001$ and recorded $100,000$ data points (ie.\
$100$ time units).

\begin{figure}[htb]
\centerline{%
\includegraphics[width=.695\textwidth]{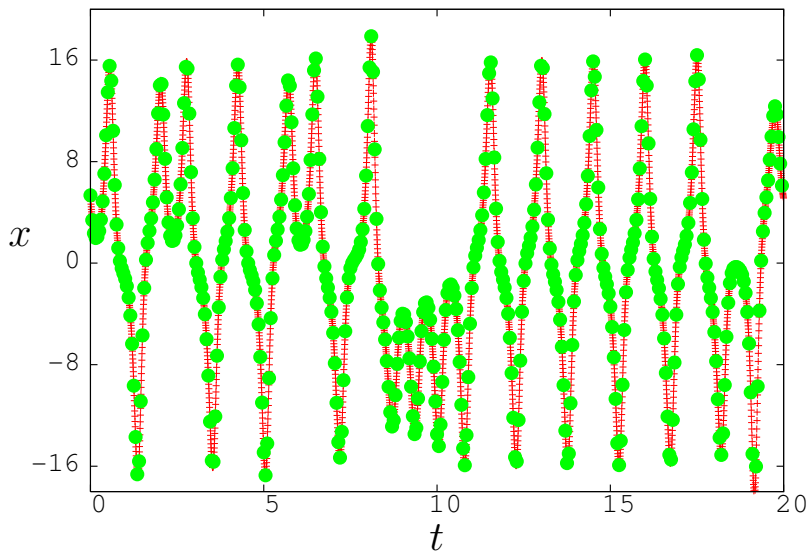}
}
\caption{Plot of the observable $\phi(t)=x(t)$ for the
Lorenz system (\ref{lorenz}).  The finely sampled data (red) are
sampled at $\tau_s=0.005$ time units. The coarsely sampled data (green
filled circles) are sampled at $\tau_s=0.05$ time units.}
\label{fig-sampleT}
\end{figure}

Fig.~\ref{fig-sampleT} shows an oversampled and a sufficiently
coarsely sampled observable for the Lorenz system (\ref{lorenz}). The
finely sampled time series ($\tau_s=0.005$) yields $K\approx0$ even
for $N=100,000$ whereas the coarsely sampled data ($\tau_s=0.05$)
yields $K\approx 1$ already for $N=5,000$.

A good choice of the sampling time $\tau_s$ can often be obtained by
visual inspection as in Fig.~\ref{fig-sampleT}. A more refined method
is to use the first minimum of
the {\em mutual information} \cite{FraserSwinney86,Kantz}. For the
data depicted in Fig.~\ref{fig-sampleT} this method yields $\tau_s=0.17$
(roughly a quarter of the oscillation period). 
Note however that in this particular instance the 
smaller sampling time $\tau_s=0.05$ already gives $K\approx 1$ and
extracts a longer time series from the data in Fig.~\ref{fig-sampleT}.
In general, the optimal sampling time will depend on the dynamical
system and the time series under consideration. We refer the reader to
\cite{Kantz} for a discussion on optimal time delays in the context of
phase space reconstruction.

Although oversampling is a practical problem for data series of finite
size, it should be emphasized that theoretically the test works for all 
sampling times $\tau_s$ in the limit $N\to \infty$.

\subsection{Oversampling and power spectra}
For continuous time systems, the mean square displacement is defined as
\begin{align*}
M_c(t) = \lim_{T\to\infty}\frac{1}{T} \int_0^T(p_c(t+\tau)-p_c(\tau))^2\, + \,
(q_c(t+\tau)-q_c(\tau))^2\; d\tau \; .
\end{align*}
For a time series sampled with sample time $\tau_s$ this can be
approximated by
\begin{align*}
M_c(n) = \lim_{N\to\infty}\frac{1}{N} \sum_{j=1}^{N} 
\left( [p_{c\tau_s}(j+n)-p_{c\tau_s}(j)]^2\, + \, [q_{c\tau_s}(j+n)-q_{c\tau_s}(j)]^2 \right)
\tau_s^2\; .
\end{align*}
Similarly the power spectrum for the time-continuous case discretizes
to
\begin{align} 
\label{eq-Sc}
S(\nu)=\lim_{n\to\infty}\frac1n
E\Bigl|\sum_{j=0}^{n-1}e^{2 \pi i\frac{\nu}{\nu_s} j}\phi(j)\Bigr|^2
\tau_s^2, 
\end{align}
where $\nu_s=1/\tau_s$ is the sample frequency. 
The power spectrum consists of discrete peaks if the underlying system
is regular, and is nowhere zero for a large class of chaotic systems
\cite{MGapp}. However, for chaotic systems the power spectrum decays for
large frequencies $\nu$, and so for frequencies larger than some
$\nu_{\rm{max}}$ the power spectrum is zero for
all practical purposes.

Comparing (\ref{eq-Sc}) with the power spectrum (\ref{eq-S}) for
discrete-time data, we identify
\[
c = 2 \pi \frac{\nu}{\nu_s},  \quad \nu \in[0,\nu_{\rm{max}}]\; .
\]
Sampling at the Nyquist rate with $\nu_s^\star=2
\nu_{\rm{max}}$ yields $c\in(0,\pi)$ as before. However,
oversampling at a higher frequency $\nu_s>\nu_s^\star$, restricts the
effective choices of $c$ to $c\in(0,c^\star)$ where
$c^\star=\frac{\nu_s^\star}{\nu_s} \pi <\pi$. 
There is now a positive probability that the test for chaos will incorrectly
yield $K=0$ since it is possible that more than half of the randomly
chosen values of $c\in(0,\pi)$ will lie in $(c^\star,\pi)$.

\begin{figure}[htb]
\centerline{%
\includegraphics[width=.495\textwidth]{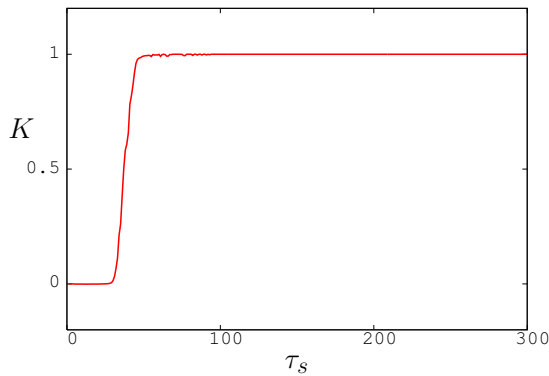}
}
\caption{Plot of $K$ as a function of the sample time $\tau_s$ for the
Lorenz system (\ref{lorenz}).
The sample time is measured in units of $\Delta t=0.001$.}
\label{fig-sampleK}
\end{figure}

\begin{figure}[htb]
\centerline{%
\includegraphics[width=.495\textwidth]{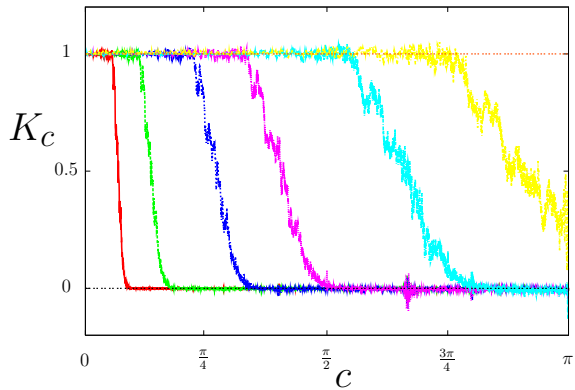}
}
\caption{Plot of $K_c$ as a function of the frequency $c$ for the
Lorenz system (\ref{lorenz}).  From left to right we used $\tau_s=5
\Delta t$, $\tau_s=10 \Delta t$, $\tau_s=20 \Delta t$, $\tau_s=30
\Delta t$, $\tau_s=50 \Delta t$, $\tau_s=70 \Delta t$. The linear
scaling of the range of $c$ for which $K_c\approx 1$ is evident in the
relative spacing of the respective lines. }
\label{fig-sampleKc}
\end{figure}

We illustrate the previous argument using the Lorenz system
(\ref{lorenz}) sampled with $\tau_s$ ranging from $\tau_s=\Delta t$ up
to $\tau_s=300\Delta t$. In Fig.~\ref{fig-sampleK} the median of the
asymptotic growth rate $K$ is shown as a function of the sample
time. For data that is too finely sampled, we obtain $K=0$ although
the dynamics is actually chaotic.
Fig.~\ref{fig-sampleKc} illustrates how 
the range of effective values of $c$ depends on the sampling time~$\tau_s$.


\section{Noise contaminated data}
\label{sec-noise}

Real-world data is invariably contaminated with noise.  Any method for
distinguishing regular from chaotic dynamics can only succeed if the
noise-level is sufficiently small.  There are various standard noise
reduction techniques \cite{Kantz} that may be applied in advance of
applying any given test for chaos.  In addition, the test itself may
be modified.  Below we indicate a modification of the $0$--$1$ test for
chaos that makes it more robust to the presence of noise.

In \cite{GM05} we introduced a version of the $0$--$1$ test that works
well for data contaminated with measurement noise. (This is the test
as presented in Sections~\ref{sec-MSQ} and~\ref{sec-Kc}, but using
$M_c(n)$ instead of $D_c(n)$ and using the regression method instead
of the correlation method.) We showed that the $0$--$1$ test clearly
outperforms tangent space methods and compares favourably to ``direct
methods'' based on phase space reconstruction. The improvements in
this paper have made our test extremely sensitive to weak chaos.
However, an unavoidable consequence is an increased sensitivity also
to noise (see Fig.~\ref{fig-noise} below).

It turns out that the success of the version of the test in
\cite{GM05} is due to the oscillatory term $V_{\rm{osc}}(c,n) =
(E\phi)^2 \frac{1-\cos nc}{1-\cos c}$ that we subtracted in
Section~\ref{sec-MSQ} to define the modified mean-square-displacement
$D_c(n)=M_c(n)-V_{\rm{osc}}(c,n)$. This term desensitizes the test and
damps the ability to detect slow growth of the
mean-square-displacement for time-series data of moderate length.
Instead of reintroducing this term we adopt a more flexible approach,
defining
\begin{align*}
D^\star_c(n)=D_c(n) + \alpha V_{\rm damp}(n), \quad 
V_{\rm damp}(n)=(E\phi)^2 \sin(\sqrt{2}n)\; .
\end{align*}
(The frequency $\sqrt{2}$ was chosen arbitrarily.)  For $\alpha$
large, we expect $K=0$. The amplitude $\alpha$ of the term $V_{\rm
damp}(n)$ controls the sensitivity of the test to weak noise and
simultaneously to weak chaos. This trade-off is unavoidable in any
test for chaos.

As an illustration, we consider the logistic map with measurement
noise.  Take as observable $\phi(n)=x_n$ and write
\[
\tilde\phi(n)=\phi(n)(1 + \frac{\epsilon}{100}\eta_n)
\]
where $\eta_n$ are i.i.d.\ random variables drawn from a uniform
distribution on $[-1,1]$ and $\epsilon$ is the noise-level in percent.
Fig.~\ref{fig-noise} shows how the undamped version of the test in
this paper copes with a noise level of $10\%$ and the improvement that
is obtained by using the damped mean-square-displacement $D_c^*(n)$.
We obtain similar results for normally distributed noise.

\begin{figure}[htb]
\centerline{%
\includegraphics[width=.495\textwidth]{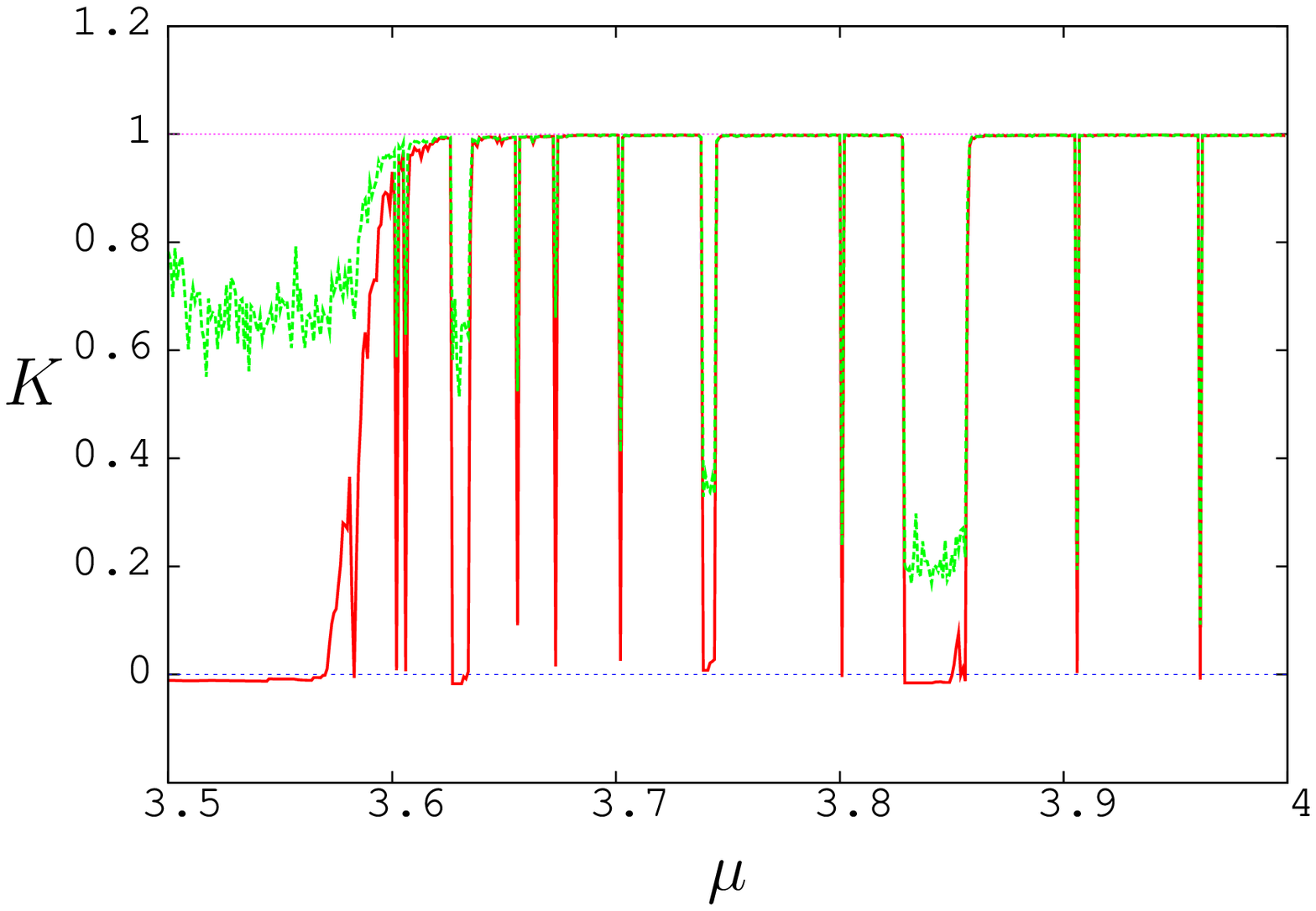}
\includegraphics[width=.495\textwidth]{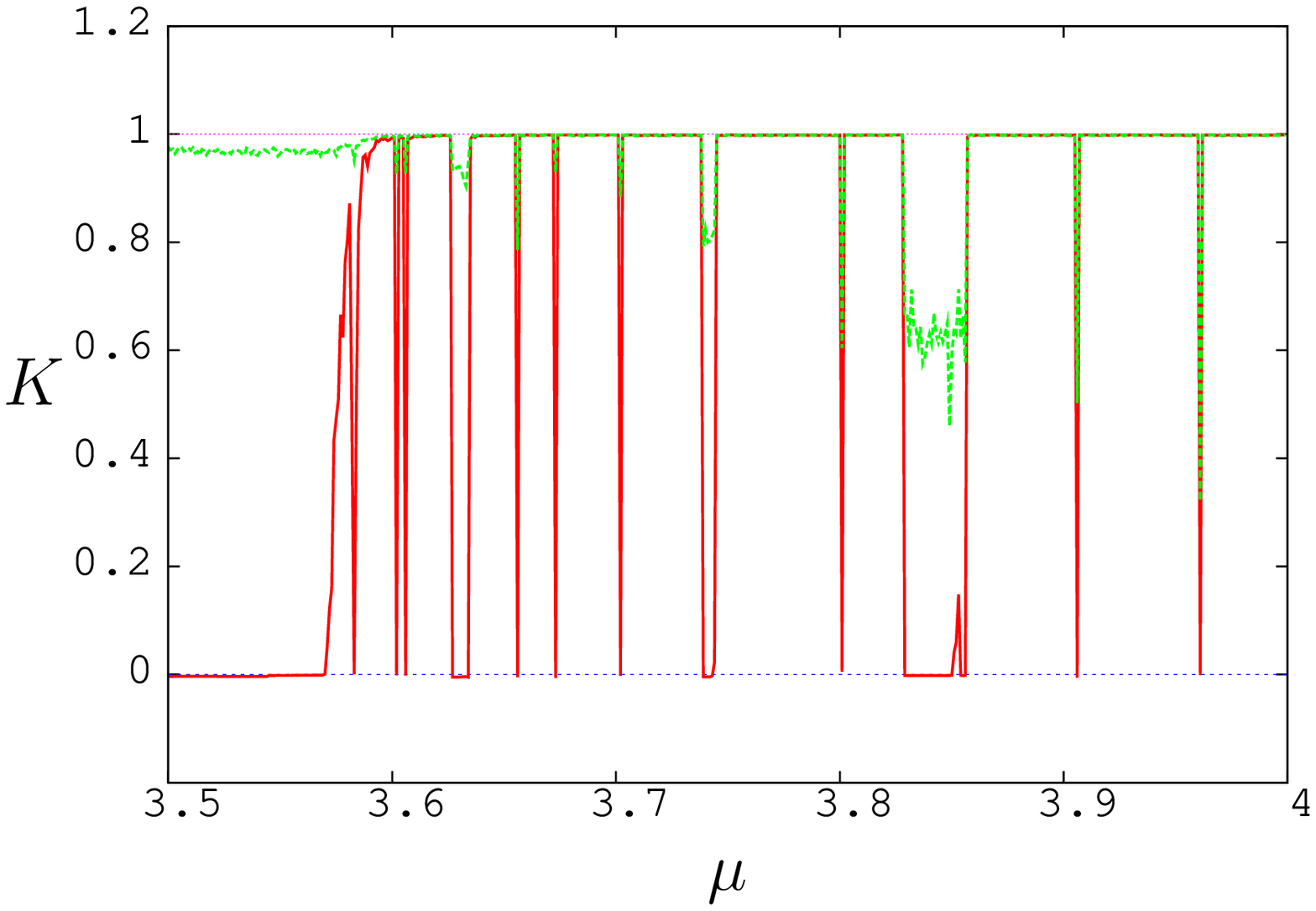}
}

\vspace{1ex}
\centerline{%
\includegraphics[width=.495\textwidth]{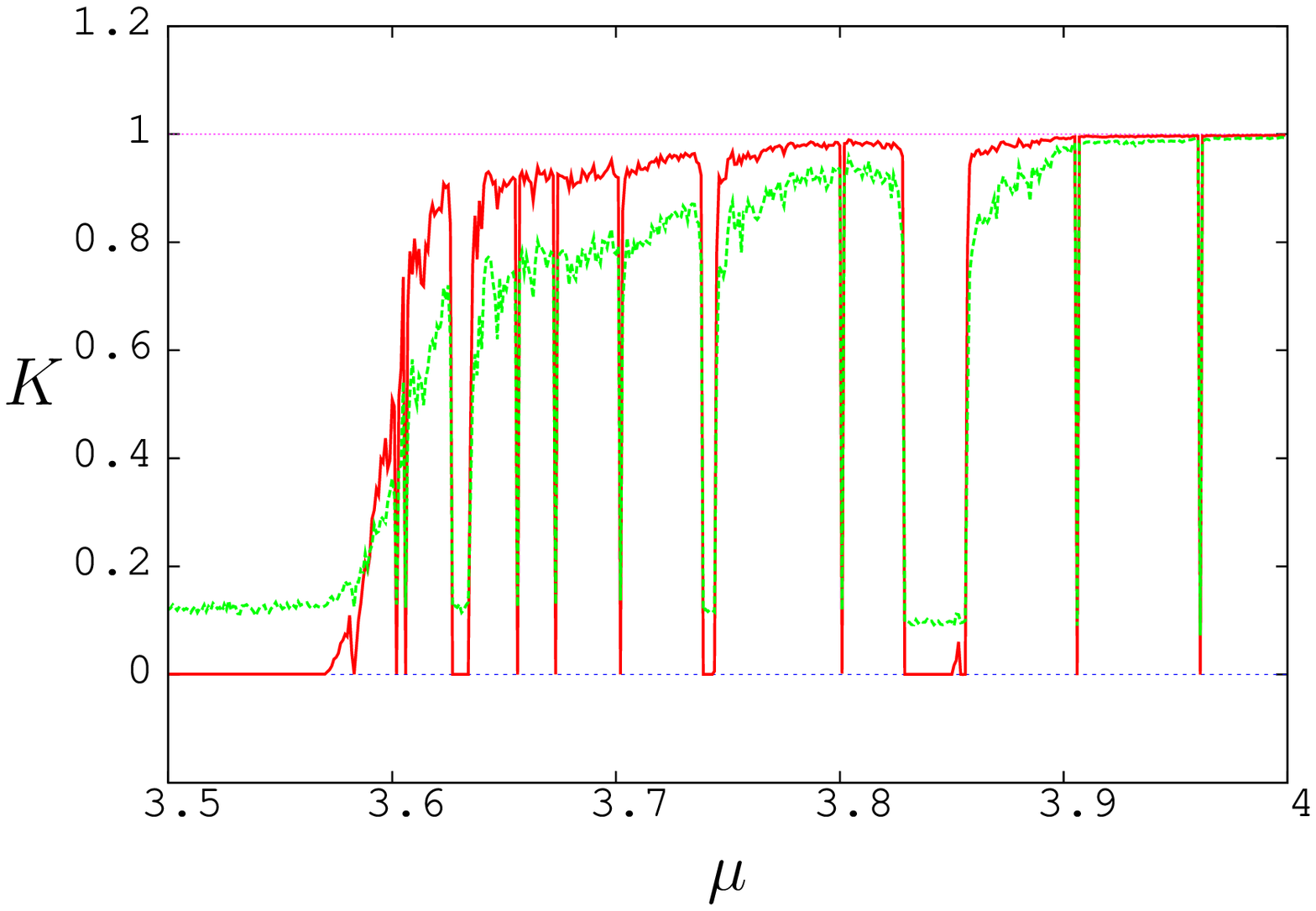}
}
\caption{
Plot of $K$ versus $\mu$ for the logistic map increased in increments
of $0.001$.  The darker (red) lines were computed using clean data.
The lighter (green) lines were computed after addition of $10\%$
uniformly distributed measurement noise.  Both lines were computed
using the undamped mean-square-displacement $D_c(n)$.  Left: $N=1000$
using the undamped mean-square-displacement $D_c(n)$, Right: $N=5000$
using $D_c(n)$.  Bottom: $N=5000$ using the damped
mean-square-displacement $D_c^*(n)$ with $\alpha=2.5$. }
\label{fig-noise}
\end{figure}

\begin{rmk}
Under the assumption that the noise is diffusive and not correlated
with the dynamics, the mean square displacement for data contaminated
with measurement noise may be written as
\[
D_c(n)=(V_{\rm{dyn}}(c)+V_{\rm{noise}}(c))n + o(n)\;,
\]
where for a given value of $c$, $V_{\rm{dyn}}(c)$ is the variance
associated with the deterministic dynamics and $V_{\rm{noise}}(c)$ the
variance associated with the measurement noise. Consider an idealized
situation where the value of $V_{\rm{noise}}(c)$ is roughly constant
as a parameter $\lambda$ is varied.  Suppose further that the dynamics
is known to be regular at $\lambda=\lambda_0$.  Then we may estimate
$V_{\rm{noise}}(c)$ by making a gauge-measurement at
$\lambda=\lambda_0$, applying the correlation method to $D_c(n)-Vn$.
The unique value $V=V_c(\lambda_0)$ which yields $K_c=0$ is our
estimate for $V_{\rm{noise}}(c)$.  For other values of $\lambda$ we
may now apply the correlation method to $D_c(n)-V_c(\lambda_0)n$.
\end{rmk}


\section{Discussion}
\label{sec-Discussion}

We have presented a guide for the implementation of the $0$--$1$ test
for chaos. At the same time, we have introduced an improved version of
the test which uses analytical expressions derived in \cite{GMprep}.
Issues such as oversampling for continuous-time data and the presence
of noise have been discussed.  We hope that this guide will be helpful
for scientists who would like to use the test.

There are numerous methods in the literature for distinguishing
between deterministic and chaotic dynamics.  In our previous
papers~\cite{GM04,GM05}, we made a careful comparison of the $0$--$1$
test with methods for computing the maximal Lyapunov exponent.
Another method is to use the power spectrum for which there are
efficient computational techniques.  It should be pointed out however
that these techniques generally rely on the Wiener-Khintchine theorem
which assumes summable decay of correlations and hence excludes
periodic and quasiperiodic dynamics.  Hence to use power spectra as a
test for chaos, it seems necessary to avoid the Wiener-Khintchine
theorem and to work directly with the expression
$\lim_{n\to\infty}\frac1n E \Bigl|\sum_{j=0}^{n-1}e^{ijc} \phi(j)
\Bigr|^2$. Equation~\eqref{eq-S} shows the relationship between the
$0$--$1$ test and power spectra, and our test can be viewed as a way of
condensing the information relevant for chaoticity or regularity
contained in the power spectrum into a single binary number.


\paragraph{Acknowledgments}
We would like to thank Ramon Xulvi-Brunet for pointing us towards the
correlation method, and for explaining us the problem of oversampling
in power spectra. We would like to thank Michael Breakspear and John
Dawes for providing encouraging feedback on an earlier version of the
manuscript. GAG is partly supported by ARC grant DP0452147 and
DP0667065. The research of IM was partly supported by EPSRC grant
EP/D055520/1 and by a Leverhulme Research Fellowship. IM is grateful
to the University of Sydney for its hospitality.


\end{document}